\documentclass[conference]{IEEEtran}
\usepackage[noadjust]{cite}
\usepackage{enumerate}
\usepackage{bm}
\usepackage{amssymb}
\usepackage{amsmath}
\usepackage{amsthm}
\usepackage{graphicx}
\usepackage{makeidx}
\usepackage{multicol}
\usepackage{comment}
\usepackage{color}
\usepackage[table,dvipsnames]{xcolor}

\usepackage{algorithm}
\usepackage{algpseudocode}
\usepackage{multirow}
\usepackage{array}
\newcolumntype{?}{!{\vrule width 0.4pt}}
\usepackage{makecell}
\usepackage{tabularx}
\usepackage{adjustbox}
\usepackage{stfloats}
\usepackage{arydshln}
\usepackage[normalem]{ulem}
\usepackage{subcaption}
\usepackage{setspace}

\algblock{ParFor}{EndParFor}
\algnewcommand\algorithmicparfor{\textbf{parFor}}
\algnewcommand\algorithmicendparfor{\textbf{end\ parFor}}
\algrenewtext{ParFor}[1]{\algorithmicparfor\ #1}
\algrenewtext{EndParFor}{\algorithmicendparfor}

\algblock{ParForAll}{EndParForAll}
\algnewcommand\algorithmicparforall{\textbf{parForAll}}
\algnewcommand\algorithmicendparforall{\textbf{end\ parForAll}}
\algrenewtext{ParForAll}[1]{\algorithmicparforall\ #1}
\algrenewtext{EndParForAll}{\algorithmicendparforall}

\definecolor{Gray}{gray}{0.3}
\theoremstyle{definition}
\newtheorem{theorem}{Theorem}
\newtheorem*{mproof}{Proof}

\newtheorem{proposition}[theorem]{Proposition}

\newtheorem{remark}{Remark}
\newtheorem{remarkone}{Remark}[remark]

\begin{document}
\title{\large\bf 
Searching and Sorting With $O(n^2)$ Processors in $O(1)$ Time
\vspace{-14pt}}
\author{
{\bf Taeyoung An {\rm and} A. Yavuz Oru\c{c}  }\\Department of Electrical Engineering\\ University of Maryland\\ College Park, MD 20742
\vspace{-15pt}}
\maketitle
\nopagebreak[4]
\setcounter{page}{1}
\begin{abstract}
The proliferation of number of processing elements (PEs) in parallel computer systems, along with the use of more extensive parallelization of algorithms causes the interprocessor communications dominate VLSI chip space. 
This paper proposes a new architecture to overcome this issue by using simple crosspoint switches to pair PEs instead of a complex interconnection network. Based on the cyclic permutation wiring idea described in~\cite{oruc2016self}, this pairing leads to a linear crosspoint array of $n(n-1)/2$ processing elements and as many crosspoints.  We demonstrate the versatility of this new parallel architecture by designing fast searching and sorting algorithms for it.
In particular, we show that finding a minimum, maximum, and searching a list of $n$ elements can all be performed in $O(1)$ time with elementary logic gates with $O(n)$ fan-in, and in $O(\lg n)$ time with $O(1)$ fan-in. We further show that sorting a list of $n$ elements can also be carried out in $O(1)$ time using elementary logic gates with $O(n)$ fan-in and threshold logic gates. The sorting time increases to $O(\lg n\lg\lg n)$ if only elementary logic gates with $O(1)$ fan-in are used.  The algorithm can find the maximum among $n$ elements in $O(1)$ time, and sort $n$ elements in $O(\lg n (\lg\lg n))$ time. In addition, we show how  other fundamental queries  can be handled within the same order of  time complexities.
\end{abstract}

\vspace{-6pt}
\section{Introduction}

\vspace{-6pt}\noindent
With the emerging  big data problems in distributed and cloud-based computing systems,  designing efficient on-chip networks for  fast searching and sorting extremely large sets of data has become a critical task. To be sure, there exist myriad searching and sorting algorithms~\cite{knuth1973art,akl2014parallel} that can be applied to processing big data in distributed and cloud-based systems, but many of these algorithms are either not sufficiently fast to deal with such large data sets or they require parallel processing elements~(PEs) with overly complex interconnection networks. In this paper, we introduce a tightly coupled linear array processing model with $O(n^2)$ simple processing elements  in which every pair of processing element  is connected by a direct link (an on-and-off switch). Unlike a conventional linear array of $O(n)$ processing elements~\cite{yen1993balancing,pan1998efficient,pan1998basic,pan1998linear,datta2002fast}, this linear array model, called a 1D-Crosspoint Array,  uses $O(n^2)$ processing elements, each of which serves as a simple compare-and-exchange operator, together with some minimal combinational circuit functions. We provide counting-based searching and sorting algorithms on this 1D-Crosspoint Array. Our search algorithm takes $O(1)$ time and sorting algorithm takes $O(\lg n\lg\lg n)$ time with constant fan-in elementary operations, and $O(1)$ time with threshold gates, making them highly competitive with  parallel searching and sorting architectures that have previously been reported in the literature. For comparison, we provide a brief survey of such architectures here. Earliest results on parallel sorting appeared in~\cite{batcher1968sorting, habermann1972parallel, baudet1975optimal,valiant1975parallelism,thompson1977sorting,preparata1978new,nassimi1979bitonic,schnorr1986optimal}.  Batcher's sorters are non-adaptive or oblivious in that they are constructed by  a set of $2\times 2$  switches connected together in stages to compare and exchange keys to sort them~\cite{batcher1968sorting}.  Batcher's odd-even and bitonic sorters use $O(n\lg^2 n)$ compare-and-exchange switches and have $O(\lg^2 n)$ sorting time. Another non-adaptive sorting network is the AKS network  that can sort a set of $n$ keys in $O(\lg n)$ time using $O(n\lg n)$ comparator/exchange switches. The main issue with the AKS network is the large constants in its hardware and sorting time complexities.  Adaptive techniques are also used in sorting as described~\cite{chien1994adaptive} for sorting binary keys in $O(\lg n)$ gate delays with $O(n)$ constant fan-in and fan-out gates.
Several other parallel realizations of sorting algorithms have also been reported in the literature. Many of these rely on a mesh-connected parallel computer model~\cite{thompson1977sorting,nassimi1979bitonic}, or more generic SIMD processors models~\cite{habermann1972parallel,baudet1975optimal}. In general, the studies on mesh-connected parallel computer models establish that $n$ keys can be sorted on a $\sqrt{n}\times \sqrt{n}$ mesh in $O(\sqrt{n})$ time with  different constants in the order of time complexity that ranges between 3 and 6. A more realistic linear array sorting model was introduced in~\cite{yen1993balancing} to sort a list of $n$ keys in $2n$ time using $O(n)$ PEs and $O(n)$ memory space.  More recent results on sorting  on parallel computers with a mesh topology make stronger claims on the time complexity of sorting, effectively reducing the time complexity of sorting to $O(\lg n\lg\lg n)$ with $n$ PEs. Examples of such results include those that appeared in~\cite{pan1998efficient,pan1998basic,pan1998linear,datta2002fast,he2009optimal}. These efforts rely on a reconfigurable pipelined bus system, called {\em  the LARPBS} (Linear Array, Reconfigurable Pipelined Bus System). The work in~\cite{datta2002fast} reduced the time complexity to $O(\lg^2\lg n)$ using $n^{1+\epsilon}$ PEs, where $0 < \epsilon < 1,$ and the result  in~\cite{he2009optimal} provided an $O(\lg n)$-time sorting algorithm on the LARPBS model with $O(n)$ PEs. The time complexity of sorting was reduced further to $O(1)$ in~\cite{lin1992sorting} using a mesh topology by assuming that PEs in such a topology may communicate with each other in $O(1)$ time. We find this assumption impractical as it ignores wire delays by assuming that bus transmissions between PEs take $O(1)$ time. 

Independent from these results, Valiant reported an abstract parallel processor model and studied the parallel time complexity of merging and sorting problems without specifying a particular topology~\cite{valiant1975parallelism}. He presented both lower and upper bounds on the parallel time complexity of merging and sorting on this abstract model. Valiant's work is important in that it guides what is feasible theoretically as the number of processing elements is varied from two PEs  to $n(n-1)/2$ PEs, even though it is not at all obvious how his merging and sorting algorithms can be mapped to an actual parallel processor architecture. In this paper, we attempt to fill this void in one particular case, namely when Valiant's abstract model assumes $n(n-1)/2$ processing elements. We introduce a realistic model for interprocessor communications by placing direct edges or on-and-off switches only between physically adjacent PEs.  More precisely, our a 1D linear array architecture with $\frac{n(n-1)}{2} + 1$ PEs  provides an $O(1)$ time sorting algorithm, matching the time complexity of sorting a list elements on Valiant's parallel processing model~\cite{valiant1975parallelism} when $O(n^2)$ PE's are used.  It should be pointed out that Valiant uses an underlying topology with $O(n)$ fan-in and fan-out to sort a list of $n$ elements. The underlying topology of our 1D-Crosspoint Array  assumes a fan-in and fan-out of $2,$ but to accomplish $O(1)$ sorting time, we employ a threshold logic circuit, which by definition requires $O(n)$ fan-in. We also employ a distribution network with $O(n)$ fan-out in our $O(1)$ time sorting algorithm. The same assumptions hold for finding the minimum and maximum of a list of elements as in Valiant's parallel processer model.
The rest of the paper is organized as follows. In the next section, we describe the 1D-Crosspoint Array and prove the minimum number of PEs needed to accomplish our searching and sorting time complexity results. In Section~\ref{sec:realization}, we explain how the new architecture can be constructed for any number of PEs. Next in Section~\ref{sec:parEnumSort}, we introduce the parallel sorting algorithm suited for the new architecture and analyze the complexity of the algorithm. We also extend this algorithm to finding the minimum, maximum of a list of elements as well as searching for an element. The paper is concluded in Section~\ref{sec:discussion}, with a discussion that includes the comparison of our sorting algorithm with some of the well-known parallel sorting algorithms.

\vspace{-7pt}
\section{The 1D-Crosspoint Array}
\label{OneDCrosspointArray}

\begin{figure}[b]
\includegraphics[width=\columnwidth]{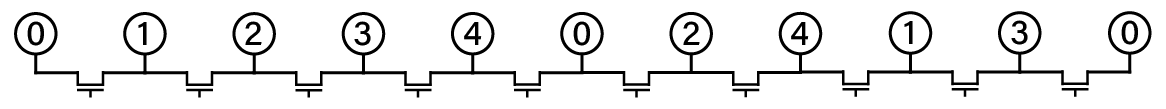}
\vspace{-18pt}
\caption{1D-Crosspoint Array for $n=5$}
\vspace{-2pt}
\label{fig:4coreEx}
\end{figure}

\vspace{-3pt}
\noindent A one dimensional (1D)-crosspoint array is the simplest form of PE pairing that establishes a baseline of layout and wiring complexity as depicted in Figure~\ref{fig:4coreEx} for a $5$-PE network. 
We say that a crosspoint array  has a 1D-layout if (a) each PE is paired with at most two pairs and (b) PEs are placed on a straight path. As seen in the figure, all $\binom{5}{2}=10$ pairings of five PEs are realizable in this layout of 10-crosspoint array in which a PE is replicated no more than three times. Each of the five PEs is replicated twice, where the copies will be referred to as replicates. 
Replicates are created to distribute the workload to multiple PEs and attain parallelism. They will be given the same input dataset, and will be carrying out the same computation with the neighboring PEs(replicates).
Thus, replicates are similar to PEs and when there is no ambiguity, they will be viewed like PEs. Each PE and its replicates form a class so that PEs and their replicates are divided into non-overlapping sets. 
Two classes will be called adjacent if they have two  elements (PEs or replicates) that share a crosspoint.  Our goal here is to have a pair of elements, i.e., PEs or replicates one from each of the $\binom{n}{2}$ pairs of $n$ classes share a crosspoint in a 1D-Crosspoint Array, using a minimum number of crosspoints and replicates and under the assumption that no PE(replicate) is adjacent to more than two PEs (replicates). 

To this end, we have the following results.
\vspace{-5pt}
\begin{proposition}
    A 1D-Crosspoint Array in which every pair of  $n$ classes is  adjacent requires at least $\binom{n}{2}$ crosspoints and $\binom{n}{2}+1$ PEs in all $n$ classes. 
    \label{prop:base}
\end{proposition}
\vspace{-12pt}
\begin{mproof}
	It is obvious that each pair of PEs requires a crosspoint to be adjacent, and hence $\binom{n}{2}$ pairs of classes require $\binom{n}{2}$ crosspoints. It is also obvious that $\binom{n}{2}$ crosspoints require $\binom{n}{2}+1$ PEs, given that no PE is connected to more than two PEs.\qed
\end{mproof}
\vspace{-12pt}
\begin{proposition}
	In a 1D-Crosspoint Array with $n$ classes, if every pair of  $n$ classes is adjacent then each class must have no fewer than $\left\lceil\frac{n-1}{2}\right\rceil$ replicates.
	\label{prop:numOfRepl}
\end{proposition}
\vspace{-10pt}
\begin{mproof}
	A PE on a 1D-Crosspoint Array can be adjacent to at most two PEs. Thus, any given class needs at least $\frac{n-1}{2}$ replicates to be adjacent to the PEs of all other $n-1$ classes. Since number of replicates can only be an integer, $\frac{n-1}{2}$ rounds up to the next larger integer, i.e., $\left\lceil\frac{n-1}{2}\right\rceil$.\qed
\end{mproof}
\vspace{-5pt}
\noindent The last proposition can be strengthened for the two PEs at the end of the 1D-Crosspoint Array:
\vspace{-5pt}
\begin{proposition}
	If the two PEs at the two ends of the 1D-Crosspoint Array belong to the same class then that class must contain at least $\lceil\frac{n+1}{2}\rceil$ PEs in order to be adjacent to $n-1$ PEs from $n-1$ different classes.
	\label{prop:endSame}
\end{proposition}
\vspace{-10pt}
\begin{mproof}
	Let's suppose that the two PEs at the two ends belong to class~$\alpha$. These two PEs can only be adjacent to two PEs in total. Therefore, for class~$\alpha$ to be adjacent to $n-1$ other classes, it must be adjacent to some $(n-1)-2=n-3$ classes if we exclude its adjacencies through the two PEs at the two ends. As in Proposition~\ref{prop:numOfRepl}, we can deduce that class $\alpha$ needs to have at least $\frac{n-3}{2}$ PEs. Hence, including the PEs at the ends, PE $\alpha$ needs at least $\frac{n-3}{2}+2=\frac{n+1}{2}$ PEs in total. Again, since the number of replicates can only be an integer, it rounds up to $\left\lceil \frac{n+1}{2} \right\rceil$. \qed
\end{mproof}
\vspace{-10pt}
\begin{proposition}
\label{prop:endDiff}
	If the two PEs at the two ends of the 1D-Crosspoint Array belong to different classes then those two classes each contain at least $\left\lceil \frac{n}{2} \right\rceil $ PEs in order to be adjacent to $n-1$ PEs from $n-1$ classes.
\end{proposition}
\vspace{-10pt}
\begin{mproof}
	This time, suppose that the PEs at the two ends of the array belong to class~$\alpha$ and $\beta$. The class~$\alpha$ can be adjacent to only one class  via the PE at its end, and similarly class~$\beta$  can be adjacent to only one class on the other end. Therefore, for class $\alpha$ to be adjacent to  $n-1$ classes, it must be adjacent to $n-2$ classes via its remaining PEs and the same holds for class~$\beta$. This implies that class~$\alpha$ and $\beta$ must each have at least $\frac{n-2}{2}=\frac{n}{2}-1$ PEs. Hence, including the PE at its end,  class $\alpha$ must contain at least $\left\lceil\frac{n}{2}-1+1\right\rceil=\left\lceil\frac{n}{2}\right\rceil$ PEs. The same holds for class~$\beta$, and hence the statement. \qed
\end{mproof}
\vspace{-7pt}
\noindent Combining Propositions~\ref{prop:base}, \ref{prop:endSame}, and \ref{prop:endDiff}, the following holds:
\vspace{-7pt}
\begin{theorem}
\label{thm:minimum}
	A 1D-Crosspoint Array in which all  $n$ classes are adjacent requires at least $\frac{n(n-1)}{2}+1$ PEs (replicates) for odd $n$, and $\frac{n^2}{2}$ PEs (replicates) for even $n$.
\end{theorem}
\vspace{-10pt}
\begin{mproof}
	When the PEs at the two ends of the array belong to the same class, by Proposition~\ref{prop:endSame}, that class would need to contain $\left\lceil \frac{n+1}{2} \right\rceil$ PEs. By Proposition~\ref{prop:numOfRepl}, each of the other $n-1$ classes would need to have $\left\lceil\frac{n-1}{2}\right\rceil$ PEs. Therefore, for the 1D-Crosspoint Array to have all $\binom{n}{2}$ pairs of $n$ different classes, it would require at least
	\vspace{2pt}\begin{equation}
		\left\lceil \frac{n+1}{2} \right\rceil \times 1 + \left\lceil\frac{n-1}{2}\right\rceil \times (n-1)\text{ PEs.}
		\label{eq:endSameTotal}
	\end{equation}
	\noindent For odd $n$ Eqn.~\ref{eq:endSameTotal}  becomes $\frac{n(n-1)}{2}+1$, and for even $n$ it reduces to $\frac{n^2}{2}+1$.\\
	When the PEs at the two ends of the array belong to different classes, by Proposition~\ref{prop:endDiff}, those classes would need to contain $\left\lceil \frac{n}{2} \right\rceil$ PEs each. By Proposition~\ref{prop:numOfRepl}, each of the other $n-2$ classes would need to have $\left\lceil\frac{n-1}{2}\right\rceil$ PEs. Therefore, for all $\binom{n}{2}$ pairs of $n$ different classes of PEs to be adjacent ,  the 1D-Crosspoint Array would require at least 
	\begin{equation}
		\left\lceil \frac{n}{2} \right\rceil\times 2 + \left\lceil\frac{n-1}{2}\right\rceil\times (n-2)\text{ PEs.}
		\label{eq:endDiffTotal}
	\end{equation}
	 If $n$ is an odd number, Eqn.~\ref{eq:endDiffTotal} reduces to $\frac{n(n-1)}{2}+2,$ and  if $n$ is an even number, it becomes $\frac{n^2}{2}.$
	 \\Therefore, for all $n$ classes to be adjacent on the 1D-Crosspoint Array, there must be $\frac{n(n-1)}{2}+1$ or more PEs for odd $n$ case, and $\frac{n^2}{2}$ or more PEs for even $n$ case.\qed
\end{mproof}
\vspace{-7pt}
\section{Construction of 1D-Crosspoint Array}
\label{sec:realization}
\vspace{-3pt}
\noindent We now present a construction for a 1D-Crosspoint Array for any $n$ number of classes using a number of PEs that matches the lower bound given in Theorem~\ref{thm:minimum}. Our construction works differently for even and odd $n$ as described below.
\vspace{-4pt}
\subsection{Even $n$}
\begin{figure*}[b]\vspace{-20pt}
	\centering
	\begin{subfigure}{\columnwidth}
		\begin{align*}
			p   &= (0,1,2,3,4,5,6,7,8,9,10,11)\\[-3pt]
			p^2 &= (0,2,4,6,8,10)\;(1,3,5,7,9,11)  \\[-3pt]
			p^3 &= (0,3,6,9)\;(1,4,7,10)\;(2,5,8,11) \\[-3pt]
			p^4 &= (0,4,8)\;(1,5,9)\;(2,6,10)\;(3,7,11) \\[-3pt]
			p^5 &= (0,5,10,3,8,1,6,11,4,9,2,7)\\[-3pt]
			p^6 &= (0,6)\;(1,7)\;(2,8)\;(3,9)\;(4,10)\;(5,11)\\[-3pt]
			p^7 &= (0,7,2,9,4,11,6,1,8,3,10,5)\\[-3pt]
			p^8 &= (0,8,4)\;(1,9,5)\;(2,10,6)\;(3,11,7)\\[-3pt]
			p^9	&= (0,9,6,3)\;(1,10,7,4)\;(2,11,8,5)\\[-3pt]
			p^{10}&= (0,10,8,6,4,2)\;(1,11,9,7,5,3)\\[-3pt]
			p^{11}&= (0,11,10,9,8,7,6,5,4,3,2,1)
		\end{align*}\vspace{-10pt}
		\setlength{\tabcolsep}{2pt}
		\caption{\begin{tabular}[t]{rl}Step 1:& The cyclic permutation group,\\& where the first element being the smallest.\end{tabular}}
	\end{subfigure}
	\begin{subfigure}{\columnwidth}\vspace{4pt}
		\begin{align*}
			Q_0 =& \;\big\{(0,1,2,3,4,5,6,7,8,9,10,11),	\\
				 & \;\;\;(0,2,4,6,8,10), (0,3,6,9), (0,4,8), \\
				 & \;\;\;(0,5,10,3,8,1,6,11,4,9,2,7), (0,6)\big\} \\
			Q_1 =& \;\big\{(1,3,5,7,9,11), (1,4,7,10), (1,5,9), (1,7)\big\} \\
			Q_2 =& \;\big\{(2,5,8,11), (2,6,10), (2,8)\big\} \\
			Q_3 =& \;\big\{(3,7,11), (3,9)\big\} \\
			Q_4 =& \;\big\{(4,10)\big\} \\
			Q_5 =& \;\big\{(5,11)\big\}
		\end{align*}\vspace{12pt}
		\setlength{\tabcolsep}{2pt}
		\caption{\begin{tabular}[t]{rl}Step 2:& All of the cycles in all of $p^j$, where $0\le j\le 6,$\\& has been partitioned into $Q_i$, $0\le i\le \frac{n}{2}-1$.\end{tabular}}	
	\end{subfigure}
	\begin{subfigure}{\textwidth}\vspace{3pt}
		\includegraphics[scale=0.5]{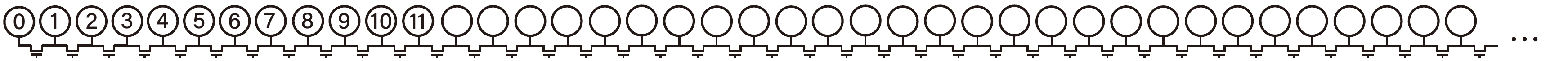}\vspace{-3pt}
		\setlength{\tabcolsep}{2pt}
		\caption{\begin{tabular}[t]{rl}Step 3:& The 1D-Crosspoint Array after placing the first cycle of $Q_0$.\\&It is assumed that $(0,1,2,3,4,5,6,7,8,9,10,11)$ was chosen.\end{tabular}}
	\end{subfigure}\\
	\begin{subfigure}{\textwidth}
		\includegraphics[scale=0.5]{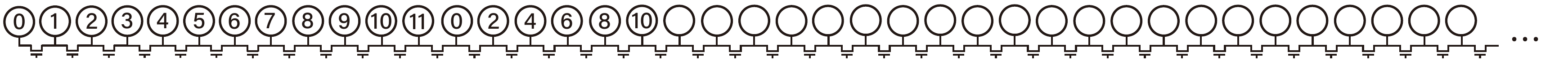}\vspace{-3pt}
		\setlength{\tabcolsep}{2pt}
		\caption{\begin{tabular}[t]{rl}Step 4:& The 1D-Crosspoint Array after placing the second cycle of $Q_0$. \\&It is assumed that $(0,2,4,8,10)$ was chosen.\end{tabular}}
		\label{fig:constructionStep4}
	\end{subfigure}\\
	\begin{subfigure}{\textwidth}
		\includegraphics[scale=0.5]{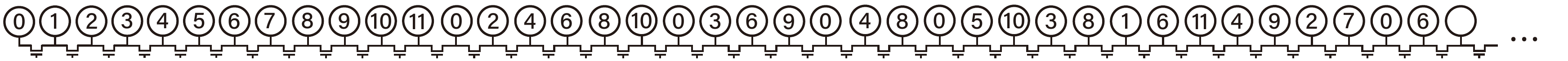}\vspace{-3pt}
		\setlength{\tabcolsep}{2pt}
		\caption{Step 5: The 1D-Crosspoint Array after placing all the cycles of $Q_0$, with the cycle $(0,6)$ from $p^\frac{n}{2}$ is placed last.}
	\end{subfigure}\\
	\begin{subfigure}{\textwidth}
		\includegraphics[scale=0.5]{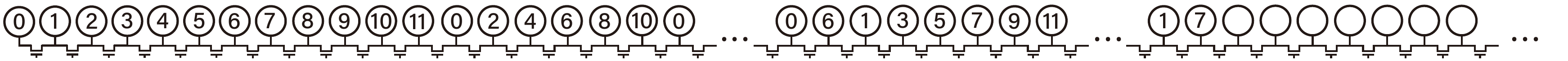}\vspace{-3pt}
		\setlength{\tabcolsep}{2pt}
		\caption{\begin{tabular}[t]{rl}Step 6:& The 1D-Crosspoint Array after placing all the cycles of $Q_1$, with the cycle $(1,7)$ from $p^\frac{n}{2}$ placed last.\\
		& Note that some of the PEs are abbreviated by `$\cdots$', due to the lack of space.\end{tabular}}
	\end{subfigure}\\
	\begin{subfigure}{\textwidth}
		\includegraphics[scale=0.5]{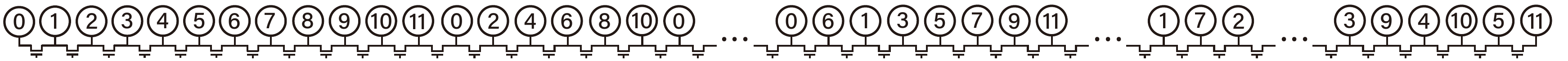}\vspace{-3pt}
		\setlength{\tabcolsep}{2pt}
		\caption{\begin{tabular}[t]{rl}Step 6:& The 1D-Crosspoint Array after placing all the cycles of every $Q_i$, where $0\le i\le \frac{n}{2}-1.$\\
		& Note that some of the PEs are abbreviated by `$\cdots$' once again due to the lack of space.\end{tabular}}
	\end{subfigure}
	\caption{An illustration of the construction algorithm for an example case of $n=12.$}\vspace{-10pt}
\label{fig:algorithm1}
\end{figure*}

\vspace{-4pt}
\noindent We borrow ideas from cyclic permutation groups as used for constructing One-sided binary tree-crossbar switch in~\cite{Oruc2015oneSideBiTree,oruc2016self}. Let $p=(0\,1\,2\,3\cdots, n-1)$ be a permutation of $n$ classes\footnote{Throughout the rest of the paper, $p$ will be fixed to this permutation.}, where $p(i) = i+1 \mod n, 0\le i\le n-1.$ We will use $p$ as the representation of layout of PEs belonging to $n$ classes, where classes whose ids are adjacent in the cycle representation of powers of $p$ will also be adjacent in the 1D-Crosspoint Array. The cyclic group of permutations generated by $p$ consists of $n$ permutations $p, p^2,p^3,\cdots, p^n$, where $p^n$ is the identity permutation. The permutation  $p^j$, where $1\le j\le  n-1$ specifies the right neighbor of class $i$ which is given by  $(i+j)\bmod{n}$. 
It is shown in~\cite{Oruc2015oneSideBiTree,oruc2016self} that $p,p^2,\cdots, p^{n-1}$ map every element to a distinct element, which can be interpreted as every PE having distinct right neighbor in the corresponding 1D-Crosspoint Array. It is also shown that $p^{n-j}$, $1\le j \le \frac{n}{2}-1$ is $p^j$ written in reverse and it represents the inverse permutation of $p^j$. The PEs are identified with the elements in $p^j$ that are generally expressed as a product of disjoint cycles. 
For example, $p$ consists of a one long cycle, i.e., a cycle of $n$ elements, $p^2$ may or may not be a long cycle depending upon $n$ being a prime or not, and so on. The elements that are adjacent in the cycles of  each $p^j$ determine the crosspoints between the PEs in some unique way. More specifically, two PEs will have a crosspoint between them if the elements that identify the two PEs are adjacent in a cycle of $p^j$ for some $j, 1\le j\le n-1.$  
Hence, the PEs are connected together using crosspoints and in~\cite{Oruc2015oneSideBiTree} PEs need not be physically adjacent when crosspoints are placed in-between them. However,  in our construction of 1D-Crosspoint Array, we restrict the placement of crosspoints between PEs that are physically adjacent.  In addition, each PE is connected to exactly one other PE in~\cite{Oruc2015oneSideBiTree}, whereas, in our work, each PE is connected to up to two other PEs. 
The last distinction we need draw between~\cite{Oruc2015oneSideBiTree} and our work is that we only use $p,p^2,\cdots,p^\frac{n}{2}$ in the construction of the 1D-Crosspoint Array even though we will make use of $p^{\frac{n}{2}+1},p^{\frac{n}{2}+2},\cdots,p^{n-1}$ in some of our proofs. 
With these ideas in mind, we now describe some preliminary facts that is used for the construction.
\begin{proposition}\vspace{-5pt}
	The number of cycles in permutation $p^j$ is equal to $\gcd(n,j),$  i.e., greatest common divisor of $n$ and $j.$
	\label{prop:numOfCycles}
\end{proposition}
\vspace{-6pt}
\begin{mproof}\vspace{-5pt}
Suppose a cycle starts with $i$. Then the following elements of the cycle are given by adding $j$ and then applying modulo $n$. The cycle ends when the result of the modulo function equals $i$ again, which is when $(i+mj)\bmod n=i$, where $m$ would be the number of elements in the cycle. This implies that the $mj$ is the least common multiple of $n$ and $j$, or $\textup{lcm}(n,j)$. Then, from the relation between $\textup{lcm}(n,j)$ and $\gcd(n,j)$, $mj=\textup{lcm}(n,j)=\frac{nj}{\gcd(n,j)}$, we get $\gcd(n,j) =\frac{n}{m},$ which is the number of cycles in $p^j.$\qed
\end{mproof}
\begin{proposition}\vspace{-5pt}
	The elements of a cycle are $\gcd(n,j)$ apart from each other. 
	\label{prop:elemKApart}
\end{proposition}
\begin{mproof}\vspace{-5pt}
	Let $k=\gcd(n,j)$ and let $u$ be an element in a cycle in permutation $p^j$, $0\le j\le \frac{n}{2}$. Further suppose $v$ is another element that belongs to the same cycle to which $u$ belongs. By the property of the cyclic permutation group, we can express $v$ as $v = (u+xj)\bmod n,$ $0\le x\le \frac{n}{k}-1,$ which implies $u+xj=an+v$, where $a$ is a non-negative integer. Since $j$ and $n$ are a multiple of $k$, letting $j=bk$ and $n=ck$, where $b$ and $c$ are positive integers, we have $u+x(bk) = a(ck)+v $ or  $v = u+(bx-ac)k. $ Now, since $v\ne u$, and $b,x,a,$ and $c$ are all integers, $(bx-ac)$ is a non-zero integer. Hence, $v$ can only be a multiple of $k$ apart from $u$, such as $u\pm k, u\pm 2k,$ etc. Therefore, every element of a cycle in $p^j$ is always $\gcd(n,j)$ apart from each other. \qed
\end{mproof}
\begin{proposition}\vspace{-5pt}
	The smallest element of any cycle in any of $p^j$, $1\le j\le \frac{n}{2}$, is less than or equal to $\frac{n}{2}-1$. 
	\label{prop:cycleMinElem}
\end{proposition}\vspace{-10pt}
\begin{mproof}
	By Proposition~\ref{prop:numOfCycles}, permutation $p^j$, $1\le j\le \frac{n}{2}$, has $\gcd(n,j)$ cycles. Suppose $p^j$ has a cycle $g_i$, where $0\le i\le\gcd(n,j)-1$. 
	Further suppose an element $m$ that is less than or equal to $\gcd(n,j)-1$ in $g_i$. Then, by Proposition~\ref{prop:elemKApart}, $m$ would be the smallest element in that cycle, since $m-\gcd(n,j)<0$. Furthermore, no other element in $g_i$ will be less than $m+\gcd(n,j)$.
	Since this holds for any $m$ that is less than $\gcd(n,j)$, all the elements less than $\gcd(n,j)$ must be distributed to distinct cycles and be the smallest element in its cycle. 
	Now, when $n$ is even and $j<n$, the maximum value of $\gcd(n,j)$  is $\frac{n}{2}$. Therefore, the smallest element in any cycle must be less than or equal to $\frac{n}{2}-1$. \qed
\end{mproof}
\vspace{-5pt}\noindent
The construction algorithm for the 1D-Crosspoint Array is shown in Algorithm~\ref{alg:evenNconstruction}, and Figure~\ref{fig:algorithm1} illustrates this algorithm for $n = 12.$
\begin{algorithm}[t]
\caption{Construction algorithm for even $n$.}
\label{alg:evenNconstruction}
\textbf{Step 1}. Suppose that the cycles in $p,p^2,\cdots,p^\frac{n}{2}$ are written so that the first element is the smallest in every one of them\footnotemark[2]. Suppose that an empty frame of a 1D-Crosspoint Array with $n^2/2$ PEs and a crosspoint between each PEs is provided without the actual assignment of class ids to the PEs.\\
\textbf{Step 2}. Partition all the cycles in all of $p,p^2,\cdots,p^\frac{n}{2}$ into $\frac{n}{2}$ sets, $Q_i$, $0\le i\le\frac{n}{2}-1$, where $Q_i$ is the set of cycles in which the first element is $i, 0\le i\le n/2-1.$ (Proposition~\ref{prop:1stElemDist} establishes that there exists such a partition of $n/2$ such sets.) Thus, $Q_0$ consists of all cycles that begin with $0$, $Q_1$ consists of all cycles that begin with $1$, and so on, 
$Q_{\frac{n}{2}-1}$ consists of all cycles that begin with $\frac{n}{2}-1$. 
Set $i=0$.\\
\textbf{Step 3}. Choose any cycle in $Q_i$ that consists of more than two elements. Then place a PE that belongs to the class $i$ into the left most PE space available in the 1D-Crosspoint Array. Next, suppose the next element in that cycle is $\alpha_i$. Place a PE that belongs to class $\alpha_i$ on the right-hand side of the previously placed PE. Repeat placing PEs in the same manner from left to right into the 1D-Crosspoint Array until all the elements in that cycle  are used.  We will refer to this process as placing a cycle on the 1D-Crosspoint Array. (See Figure~\ref{fig:constructionStep4}.)\\
\textbf{Step 4}. Choose a cycle in $Q_i$ that has not yet been chosen. Then place the newly chosen cycle into the 1D-Crosspoint Array in the same way as in Step 3. The first PE associated with the new cycle should be placed on the right-hand side of the last PE from Step 3 without skipping any PE spaces. 
\textbf{Step 5}. Repeat Step 4 until all the cycles consisting of more than two elements have been processed. Then choose the cycle in $Q_i$ that consists of only two elements (Every $Q_i$ has exactly one cycle of two elements as would be implied by Proposition~\ref{prop:twoElemCycle} and the definition of $Q_i$), and place it into the 1D-Crosspoint Array as before.\\
\textbf{Step 6}. Set $i=i+1$ and repeat  Steps 3,4, and 5 until $i=\frac{n}{2}-1.$ Again, PE spaces must not be skipped between the steps.\vspace{1pt}\\
\begin{footnotesize}
	$~~~~^2$This is done for notational convenience only.
\end{footnotesize}
\end{algorithm}
The following propositions are the propositions mentioned in Step 2 and Step 5.
\begin{proposition}\vspace{-5pt}
	The cycles in all of $p,p^2,\cdots,p^\frac{n}{2}$ can be partitioned into $n/2$ subsets $Q_i$, $0\le i\le \frac{n}{2}-1, $ where $Q_i$ is defined in Algorithm~\ref{alg:evenNconstruction}. 
	\label{prop:1stElemDist}
\end{proposition}
\begin{mproof}\vspace{-8pt}
	By Proposition~\ref{prop:cycleMinElem}, the first element of any cycle in any of $p^j$, $1\le j\le \frac{n}{2}$, is less than or equal to $\frac{n}{2}-1$. Therefore, the cycles in all of $p^j$ can be partitioned into $\frac{n}{2}$ subsets $Q_i$ as defined in Algorithm 1. \qed
\end{mproof}
\begin{proposition}
	\vspace{-5pt}
	For an even $n$, the permutation $p^\frac{n}{2}$ is the only permutation among $p^j$, where $1\le j\le\frac{n}{2}$, whose $\frac{n}{2}$ cycles are composed of two elements each.
	\label{prop:twoElemCycle}
\end{proposition} 
\begin{mproof}
	\vspace{-8pt}
	By Proposition~\ref{prop:numOfCycles}, the number of cycles in $p^j$ is $\gcd(n,j).$ When $n$ is  even and $j<n$, the maximum value of $\gcd(n,j)$  is $\frac{n}{2}$, which occurs only if $j=\frac{n}{2}$. \qed
\end{mproof}

\vspace{-3pt}\noindent
We next establish that Steps 3 through 6 construct a 1D-Crosspoint Array in which all $\binom{n}{2}$ pairs of classes of PEs are connected together by crosspoints.
Suppose that  $p^{j} = g_1g_2...g_k$, where each $g_i$ represents a cycle, and further suppose that $g_i=(b_{i,0}b_{i,1}...b_{i,r_i-1})$, where $r_i$ is the length of $g_i$. Then $p^{n-j}=g_1^{-1}g_2^{-1}...g_k^{-1}$, and we write $g_i^{-1}$ as $g_i^{-1}=(b_{i,0}b_{i,r_i-1} b_{i,r_i-2}... b_{i,1})$. 
Then, $g_i$ represent the pairs $(b_{i,0},b_{i,1}), (b_{i,1},b_{i,2}), (b_{i,2},b_{i,3}),$ $(b_{i,3},b_{i,4}),\cdots,$ $(b_{i,r_i-3},b_{i,r_i-2}),$ $(b_{i,r_i-2},b_{i,r_i-1})$ in the 1D-Crosspoint Array. 
If $r_i$ is an even number, we note that of these $r_{i-1}$ pairs,  $(b_{i,0},b_{i,1}),$ $(b_{i,2},b_{i,3}),$$\cdots,$$(b_{i,r_i-2},b_{i,r_i-1})$ will be pairs in $p^j$, and $(b_{i,1},b_{i,2}),$ $(b_{i,3},b_{i,4}),\cdots,$ $(b_{i,r_i-3},b_{i,r_i-2})$ will be pairs in $p^{n-j}$ in the one-sided binary tree-crossbar switch that is described in~\cite{Oruc2015oneSideBiTree}. 
Thus, there is a one-to-one correspondence between the pairs that the cyclic permutation group forms in 1D-Crosspoint Array and in the one-sided binary tree-crossbar switch. However, the pair $(b_{i,r_i-1}\,b_{i,0})$ that is formed in $p^{n-j}$ in the one-sided binary tree-crossbar is left out in this one-to-one correspondence. This pair is accounted for by Step 4 in Algorithm~\ref{alg:evenNconstruction}, 
by inserting a crosspoint between the last PE placed by the previously chosen cycle and very first PE placed by the newly chosen cycle, since every cycle in the same $Q_i$, $0\le i\le \frac{n}{2}-1$, start with the same $b_{i,0}$ by Steps 1 and 2.
Step 4 excludes choosing a cycle that consists only two elements, and saves it for Step 5. This is because for cycles consisting only two elements, no pair will be omitted in Steps 3 and 4, since $(b_{i,r_i-1}\,b_{i,0})=(b_{i,1}\,b_{i,0})$, which eliminates the need for placing the PE belonging to the same class as the starting PE in the same $Q_i$.\\
\noindent Now, it was proven in \cite{Oruc2015oneSideBiTree} that every $\binom{n}{2}$ pair is formed by $p,p^2,\cdots,p^{n-1}$ in the one-sided binary tree-crossbar switch. Therefore,  following steps 3 through 6,  keeping in mind the one-to-one correspondence that has been described, all $\binom{n}{2}$ pairs that  are extracted from the first $\frac{n}{2}$ permutations, $p,p^2,\cdots,p^\frac{n}{2},$ are placed in the 1D-Crosspoint Array  . \\

\vspace{-12pt}\noindent 
In the case that $r_i$ is odd, $g_i$ represents the pairs $(b_{i,0},b_{i,1}), (b_{i,1},b_{i,2}), \cdots, (b_{i,r_i-3},b_{i,r_i-2}),$ $(b_{i,r_i-2},b_{i,r_i-1})$ in 1D-Crosspoint Array. Of these $r_i-1$ pairs, $(b_{i,0},b_{i,1}),$ $(b_{i,2},b_{i,3}),$$\cdots,$$(b_{i,r_i-3},b_{i,r_i-2})$ will be pairs in $p^j$, and $(b_{i,1},b_{i,2}),$ $(b_{i,3},b_{i,4}),\cdots,$ $(b_{i,r_i-2},b_{i,r_i-1})$ will be pairs in $p^{n-j}$ in the one-sided binary tree-crossbar switch. The omitted pair is the same $(b_{i,r_i-1}\,b_{i,0})$ pair, and therefore steps 3 through 6 ensure all the cycles in all of $p,p^2,\cdots,p^\frac{n}{2}$ places all $\binom{n}{2}$ pairs in  an 1D-Crosspoint Array with crosspoints in-between them for odd $r_i$ as well.
Note that the number of elements in $p,p^2,\cdots,p^\frac{n}{2}$ is $n\times \frac{n}{2}=\frac{n^2}{2}$, which matches the lower bound in Theorem~\ref{thm:minimum}. Therefore, the 1D-Crosspoint Array described in the algorithm is optimal with respect to the number of PEs used.
However, it is important to note that the optimal lower bound in Theorem~\ref{thm:minimum} for even $n$ is greater than the minimum number of PEs in Proposition~\ref{prop:base}. 
More specifically, it is greater by $\frac{n}{2}-1$, which indicates the existence of $\frac{n}{2}-1$ redundant adjacencies on the 1D-Crosspoint Array. In fact, the pairs made of the last PE placed in step 5 and the first PE placed in the next iteration of step 3, i.e., the pairs made by elements from two different $Q_i$'s, are already created elsewhere in the 1D-Crosspoint Array.
By Proposition~\ref{prop:1stElemDist}, there are $\frac{n}{2}$ of $Q_i$'s and therefore there will be $\frac{n}{2}-1$ pairs between those $Q_i$'s which confirms the $\frac{n}{2}-1$ of redundant adjacency. We state this formally in the following remark, because it will play a key role in the odd $n$ 1D-Crosspoint Array construction and in the sorting algorithm that is presented in the next section. 
\setcounter{remark}{1}\vspace{-5pt}
\begin{remarkone}\label{remk:evenNadj}
For all even $n$, there are $\frac{n}{2}-1$ redundant adjacencies on the 1D-Crosspoint Array. \qed 
\end{remarkone}
\vspace{-10pt}
\subsection{Odd $n$}
\begin{figure}[h]
	\vspace{-2pt}
	\begin{subfigure}{\columnwidth}
		\begin{align*}
			p   &= (0,1,2,3,4,5)\\[-3pt]
			p^2 &= (0,2,4)\;(1,3,5)  \\[-3pt]
			p^3 &= (0,3)\;(1,4)\;(2,5)
		\end{align*}\vspace{-10pt}
		\caption{The cyclic permutation for $n-1=6$ case.}
	\end{subfigure}
	\begin{subfigure}{\columnwidth}
		\begin{align*}
			Q_0 =& \;\big\{(0,1,2,3,4,5), (0,2,4), (0,3)\big\} \\
			Q_1 =& \;\big\{(1,3,5), (1,4)\big\} \\
			Q_2 =& \;\big\{(2,5)\big\}
		\end{align*}\vspace{-10pt}
		\caption{All the cycles has been partitioned into $Q_i$, $0\le i\le \frac{n-1}{2}-1$.}	
	\end{subfigure}
	\begin{subfigure}{\columnwidth}
		\includegraphics[scale=0.165]{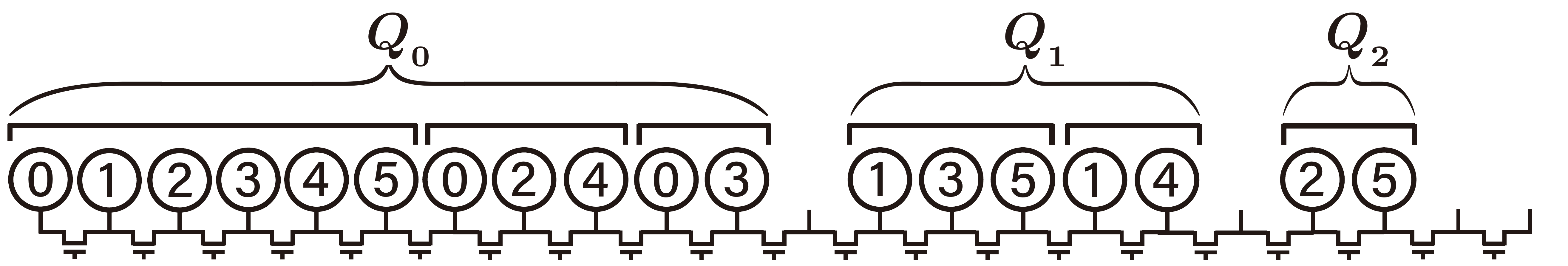}\vspace{-5pt}
		\caption{1D-Crosspoint Array for $n-1=6$. Note that a PE space was skipped between consecutive $Q_i$'s.}
		\label{fig:ex6n-1}
	\end{subfigure}
	\begin{subfigure}{\columnwidth}
		\vspace{2pt}
		\includegraphics[scale=0.165]{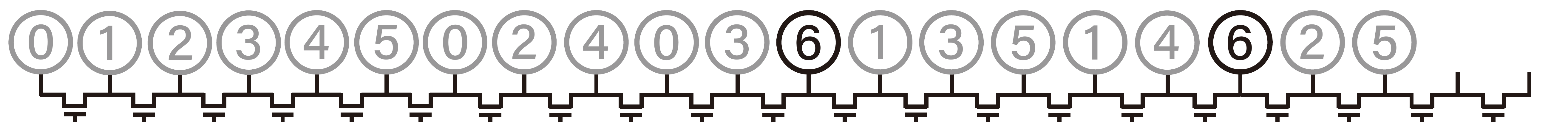}\vspace{-5pt}	
	\caption{The PE of class $n-1$ is placed in the skipped PE spaces.}
		\label{fig:ex6n1}
	\end{subfigure}
	\begin{subfigure}{\columnwidth}
		\vspace{2pt}
		\includegraphics[scale=0.165]{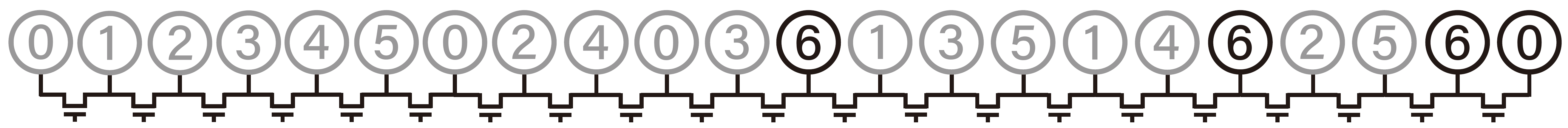}\vspace{-5pt}	
	\caption{The PE of class $n-1$ and $0$ are placed at the end.}
		\label{fig:ex6n2}
	\end{subfigure}\vspace{-3pt}
	\caption{Construction of 1D-Crosspoint Array for $n=7$ case.}
	\label{fig:ex7}
\end{figure}

\vspace{-5pt}\noindent
For an odd $n$, we first construct a 1D-Crosspoint Array for $n-1$ classes, using Algorithm~\ref{alg:evenNconstruction}, except that in Step 1, we start with an empty frame with $\frac{n(n-1)}{2}+1$ PEs, and in Step 6, we skip a PE space between the placements of consecutive $Q_i$'s, as shown in Figure~\ref{fig:ex6n-1}. When 1D-Crosspoint Array for $n-1$ classes are constructed, we place the PEs of the $n^{th}$ class, or class $n-1$, in the skipped PE spaces after Step 6, as shown in Figure~\ref{fig:ex6n1}. There will always be two empty PE spaces at the end, and we place the PE of class $n-1$ in the first empty PE space, and place the PE of class $0$ in the second one as shown in Figure~\ref{fig:ex6n2}.\\ 
It still remains to establish that all $\binom{n}{2}$ adjacencies of $n$ classes are included in the 1D-Crosspoint Array. By constructing a 1D-Crosspoint Array for $n-1$ classes, we capture every adjacencies between $n-1$ classes, and we would only need to capture adjacencies between class $n-1$ and the other classes.
Previously, in Remark~\ref{remk:evenNadj}, we pointed out the existence of redundant pairs in the 1D-Crosspoint Array, which were made by elements from two different $Q_i$'s.
By inserting class $n-1$ in between the classes that form such pairs, we capture the adjacencies between class $n-1$ and other classes without destroying any of the existing pairs. 
In fact, this procedure will ensure the 1D-Crosspoint Array to have the adjacencies between class $n-1$ and class $1,2,\cdots,n-3$. 
This is because the PE on the righthand side of the skipped PE space will be the first PE in each of the cycles in $Q_i$, $1\le i\le \frac{n-1}{2}-1$, i.e., PEs of class $1,2,\cdots,\frac{n-1}{2}-1$, and the PE on the lefthand side of the skipped PE space will be the second PE of the cycles in $p^\frac{n-1}{2}$ except the one cycle belonging to $Q_{\frac{n}{2}-1}$, i.e., PEs of class $\frac{n-1}{2},\frac{n-1}{2}+1,\cdots, n-3$.
Therefore, by placing the PE of class $n-1$ in the skipped PE spaces  we capture the adjacencies between class $n-1$ and class $1,2,\cdots,n-3$. 
The only adjacencies that are not yet captured on the 1D-Crosspoint Array are of pairs $(n-1,0)$ and $(n-1,n-2)$. These two adjacencies can be captured by placing PE of class $n-1$ and $0$ at the end of the 1D-Crosspoint Array. An example case of $n=7$ case is illustrated in Figure~\ref{fig:ex7}. 
\\Note that the number of PEs we have placed after constructing a 1D-Crosspoint Array for $n-1$ is $\frac{n-1}{2}-1+2=\frac{n+1}{2}$. Therefore, the total number of PEs is $\frac{n+1}{2}+\frac{(n-1)^2}{2}=\frac{n(n-1)}{2}+1$, which matches the minimum number of PEs needed for odd $n$ from Theorem~\ref{thm:minimum} and Proposition~\ref{prop:base}. Therefore, we can construct a 1D-Crosspoint Array for odd $n,$ where every pair of $n$ classes is adjacent only once on the 1D-Crosspoint Array.
We state this formally in the following remark as it will play a key role in the memory requirements of the sorting algorithm that is presented in the next section.
\vspace{-5pt}\begin{remarkone}\label{remk:oddNadj}
For all odd $n$ each class of PEs is adjacent to every other class of PEs exactly once.\qed
\end{remarkone}
\vspace{-10pt}
\section{Parallel Enumeration Sort}
\begin{figure*}[t!]
	\vspace{-25pt}
	\setlength\arraycolsep{2.5pt}
	\begin{center}
	\begin{tabular}{rcc|c|c|c|c|c}
	&& $C_{i,j}$ & \gape{\makecell{$C_{i_l,j_l}$\\$C_{i_r,j_r}$}}
	& \texttt{parFor} in line~\ref{line:sortCidCompL}, \ref{line:sortCidCompR}
	& \texttt{parFor} in line~\ref{line:sortElemCompL}, \ref{line:sortElemCompR}
	& \makecell{Array $T$\\after line~\ref{line:sortCompForEnd}}
	& \makecell{\texttt{parFor}\\in line~\ref{line:sortRankStore}}\\ \Xhline{0.8pt}
 
	\multirow{32}{5pt}[-15pt]{\includegraphics[scale=0.9]{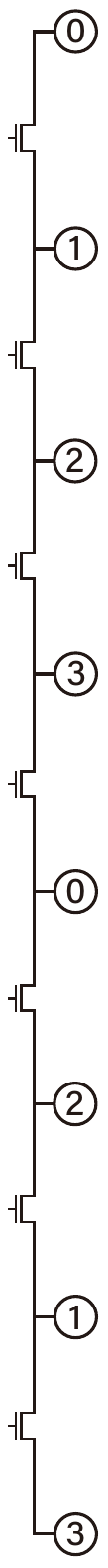}}
	&& \multirow{4}{*}{\makecell{\\$C_{0,0}$\\{\footnotesize$A[0]=6$}}}
	&&\vspace{2pt}&
	& \multirow{4}{*}{\gape{$T=\left[\color{gray}
		\begin{matrix} 0&0&0&1 \\ 1&0&0&1 \\ 1&1&0&1 \\ 0&0&0&0\\ \end{matrix}\color{black}\right] $}}
	& \multirow{4}{*}{\Gape[4pt][0pt]{$R[0]=1$}}\\  

	&&&&&&& \\ \cdashline{4-6}

	&&&\multirow{2}{*}{$C_{1,0}$}
	&\multicolumn{1}{l|}{\Gape[4pt][0pt]{$i_r=1\nless i=0$}}
	&\multicolumn{1}{l|}{$A[1]=7\nless A[0]=6$}
	&& \\

	&&&
	& \multicolumn{1}{r|}{\Gape[0pt][3pt]{$\therefore$ Receive $A[1]$ from right}}
	& \multicolumn{1}{r|}{\Gape[0pt][3pt]{$\therefore$ Send $1$ to right}}
	&& \\ \cline{3-8}
 
	&& \multirow{4}{*}{\makecell{\\$C_{1,0}$\\{\footnotesize$A[1]=7$}}}
	& \multirow{2}{*}{$C_{0,0}$}
	& \multicolumn{1}{l|}{\Gape[4pt][0pt]{$i_l=0  <   i=1$}}
	& \multicolumn{1}{l|}{Receive $1$ from left}
	& \multirow{4}{*}{\gape{$T=\left[\color{gray}
		\begin{matrix} 0&0&0&1 \\ \textcolor{black}{\textbf{1}}&0&0&1 \\ 1&1&0&1 \\ 0&0&0&0\\ \end{matrix}
		\color{black}\right] $}}
	& \multirow{4}{*}{\Gape[4pt][0pt]{$R[1]=2$}}
	\\ 

	&&&
	& \multicolumn{1}{r|}{\Gape[0pt][3pt]{$\therefore$ Send $A[1]$ to left}}
	& \multicolumn{1}{r|}{$\therefore T[1][0]\leftarrow 1$}
	&& 
	\\ \cdashline{4-6}

	&& &\multirow{2}{*}{$C_{2,0}$}
	& \multicolumn{1}{l|}{\Gape[4pt][0pt]{$i_r=2\nless i=1$}}
	&\multicolumn{1}{l|}{$A[2]=8\nless A[1]=7$}
	&&
	\\

	&&&
	& \multicolumn{1}{r|}{\Gape[0pt][3pt]{$\therefore$ Receive $A[2]$ from right}}
	&\multicolumn{1}{r|}{$\therefore$ Send $1$ to right} 
	&& \\ \cline{3-8} 
 
	&& \multirow{4}{*}{\makecell{\\$C_{2,0}$\\{\footnotesize$A[2]=8$}}}
	& \multirow{2}{*}{$C_{1,0}$}
	& \multicolumn{1}{l|}{\Gape[4pt][0pt]{$i_l=1  <   i=2$}}
	& \multicolumn{1}{l|}{Receive $1$ from left}
	& \multirow{4}{*}{\gape{$T=\left[\color{gray}
		\begin{matrix} 0&0&0&1 \\ 1&0&0&1 \\ 1&\textcolor{black}{\textbf{\underline{1}}}&0&\textcolor{black}{\textbf{1}} \\ 0&0&0&0\\ \end{matrix}
		\color{black}\right] $}}
	& \multirow{4}{*}{\Gape[4pt][0pt]{$R[2]=3$}}\\

	&&&
	& \multicolumn{1}{r|}{\Gape[0pt][3pt]{$\therefore$ Send $A[2]$ to left}}
	& \multicolumn{1}{r|}{$\therefore T[2][1]\leftarrow 1$}
	&& 
	\\ \cdashline{4-6}

	&& &\multirow{2}{*}{$C_{3,0}$}
	& \multicolumn{1}{l|}{\Gape[4pt][0pt]{$i_r=3\nless i=2$}}
	& \multicolumn{1}{l|}{$A[3]=5 < A[2]=8$}
	&& 
	\\

	&&&
	& \multicolumn{1}{r|}{\Gape[0pt][3pt]{$\therefore$ Receive $A[3]$ from right} }
	&\multicolumn{1}{r|}{$\therefore T[2][3]\leftarrow 1$, Send $0$ to right}
	&& \\ \cline{3-8} 
 
	&& \multirow{4}{*}{\makecell{\\$C_{3,0}$\\{\footnotesize$A[3]=5$}}}
	& \multirow{2}{*}{$C_{2,0}$}
	& \multicolumn{1}{l|}{\Gape[4pt][0pt]{$i_l=2<i=3$}}
	& \multirow{2}{*}{\Gape[4pt][0pt]{Receive $0$ from left}}
	& \multirow{4}{*}{\gape{$T=\left[\color{gray}
		\begin{matrix} 0&0&0&1 \\ 1&0&0&1 \\ 1&1&0&1 \\ 0&0&0&0\\ \end{matrix}
		\color{black}\right] $}}
	& \multirow{4}{*}{\Gape[4pt][0pt]{$R[3]=0$}}\\

	&&&
	& \multicolumn{1}{r|}{\Gape[0pt][3pt]{$\therefore$ Send $A[3]$ to left} }
	& 
	&& 
	\\ \cdashline{4-6}

	&& &\multirow{2}{*}{$C_{0,1}$}
	& \multicolumn{1}{l|}{\Gape[4pt][0pt]{$i_r=0   < i=3$}}
	& \multirow{2}{*}{\Gape[4pt][0pt]{Receive $0$ from right}}
	&& 
	\\

	&&&
	& \multicolumn{1}{r|}{\Gape[0pt][3pt]{$\therefore$ Send $A[3]$ to right} }
	&&& \\ \cline{3-8} 
 
	&& \multirow{4}{*}{\makecell{\\$C_{0,1}$\\{\footnotesize$A[0]=6$}}}
	& \multirow{2}{*}{$C_{3,0}$}
	& \multicolumn{1}{l|}{\Gape[4pt][0pt]{$i_l=3\nless i=0$}}
	& \multicolumn{1}{l|}{$A[3]=5 < A[0]=6$}
	& \multirow{4}{*}{\gape{$T=\left[\color{gray}
		\begin{matrix} 0&0&0&\textcolor{black}{\textbf{1}} \\ 1&0&0&1 \\ 1&1&0&1 \\ 0&0&0&0\\ \end{matrix}
		\color{black}\right] $}}
	&\\ 

	&&&
	& \multicolumn{1}{r|}{\Gape[0pt][3pt]{$\therefore$ Receive $A[3]$ from left}}
	& \multicolumn{1}{r|}{$\therefore T[0][3]\leftarrow1,$ Send $0$ to left}
	&& \\ \cdashline{4-6}

	&& &\multirow{2}{*}{$C_{2,1}$}
	& \multicolumn{1}{l|}{\Gape[4pt][0pt]{$i_r=2\nless i=0$}}
	& \multicolumn{1}{l|}{$A[2]=8\nless A[0]=6$}
	&& \\

	&&&
	& \multicolumn{1}{r|}{\Gape[0pt][3pt]{$\therefore$ Receive $A[2]$ from right}}
	& \multicolumn{1}{r|}{$\therefore$ Send $1$ to right} 
	&& \\ \cline{3-8}
 
	&& \multirow{4}{*}{\makecell{\\$C_{2,1}$\\{\footnotesize$A[2]=8$}}}
	& \multirow{2}{*}{$C_{0,1}$}
	& \multicolumn{1}{l|}{\Gape[4pt][0pt]{$i_l=0   <  i=2$}}
	& \multicolumn{1}{l|}{Receive $1$ from right}
	& \multirow{4}{*}{\gape{$T=\left[\color{gray}
		\begin{matrix} 0&0&0&1 \\ 1&0&0&1 \\ \textcolor{black}{\textbf{1}}&\textcolor{black}{\textbf{\underline{1}}}&0&1 \\ 0&0&0&0\\ \end{matrix}
		\color{black}\right] $}}
	&\\ 

	&&&
	& \multicolumn{1}{r|}{\Gape[0pt][3pt]{$\therefore$ Send $A[2]$ to left} }
	& \multicolumn{1}{r|}{$\therefore T[2][0]\leftarrow1$}
	&& \\ \cdashline{4-6}

	&& &\multirow{2}{*}{\Gape[4pt][0pt]{$C_{1,1}$}}
	& \multicolumn{1}{l|}{$i_r=1   <  i=2$}
	& \multicolumn{1}{l|}{\Gape[3pt][0pt]{Receive $1$ from right}}
	&& \\

	&&&
	& \multicolumn{1}{r|}{\Gape[0pt][3pt]{$\therefore$ Send $A[2]$ to right}}
	& \multicolumn{1}{r|}{$\therefore T[2][1]\leftarrow1$}
	&& \\ \cline{3-8} 
 
	&& \multirow{4}{*}{\makecell{\\$C_{1,1}$\\{\footnotesize$A[1]=7$}}}
	& \multirow{2}{*}{$C_{2,1}$}
	& \multicolumn{1}{l|}{\Gape[4pt][0pt]{$i_l=2\nless i=1$}}
	& \multicolumn{1}{l|}{$A[2]=8\nless A[1]=7$}
	& \multirow{4}{*}{\gape{$T=\left[\color{gray}
		\begin{matrix} 0&0&0&1 \\ 1&0&0&\textcolor{black}{\textbf{1}} \\ 1&1&0&1 \\ 0&0&0&0\\ \end{matrix}
		\color{black}\right] $}}
	&\\ 

	&&&
	& \multicolumn{1}{r|}{\Gape[0pt][3pt]{$\therefore$ Receive $A[2]$ from left} }
	& \multicolumn{1}{r|}{$\therefore$ Send $1$ to left}
	&& \\ \cdashline{4-6}

	&& &\multirow{2}{*}{$C_{3,1}$}
	& \multicolumn{1}{l|}{\Gape[4pt][0pt]{$i_r=3\nless i=1$}}
	& \multicolumn{1}{l|}{$A[3]=6 < A[1]=7$}
	&& \\

	&&&
	& \multicolumn{1}{r|}{\Gape[0pt][3pt]{$\therefore$ Receive $A[3]$ from right}}
	& \multicolumn{1}{r|}{$\therefore T[1][3]\leftarrow 1$, Send $0$ to right}
	&& \\ \cline{3-8}
 
	&& \multirow{4}{*}{\makecell{\\$C_{3,1}$\\{\footnotesize$A[3]=5$}}}
	& \multirow{2}{*}{$C_{1,1}$}
	&\multicolumn{1}{l|}{\Gape[4pt][0pt]{$i_l=1< i=3$}}
	& \multirow{2}{*}{Receive $0$ from left}
	& \multirow{4}{*}{{$T=\left[\color{gray}
		\begin{matrix} 0&0&0&1 \\ 1&0&0&1 \\ 1&1&0&1 \\ 0&0&0&0\\ \end{matrix}
		\color{black}\right] $}}
	&\\ 

	&&&
	&\multicolumn{1}{r|}{\Gape[0pt][3pt]{$\therefore$ Send $A[3]$ to left} }
	&&&\\ \cdashline{4-6}

	&&&&&\Gape[6pt][0pt]{}&& \\

	&&&&&\Gape[0pt][4pt]{}&& \\ \hline
 
	\end{tabular}
	\vspace{-3pt}
	\caption{Illustration of sorting $n=4$ elements using 1D-Crosspoint Array, with input data $A=\{6,7,8,5\}$. It is showing what is done in each \texttt{ParFor} loop.}
	\label{fig:4ex}
	\end{center}
	\vspace{-25pt}
\end{figure*}

\begin{figure*}[t!]
	\vspace{-15pt}
	\setlength\arraycolsep{2pt}
	\begin{center}
	\begin{tabular}{rcc|c|c|c|c|c}
	&& $C_{i,j}$ & \gape{\makecell{$C_{i_l,j_l}$\\$C_{i_r,j_r}$}}
	& \texttt{parFor} in line~\ref{line:sortCidCompL}, \ref{line:sortCidCompR}
	& \texttt{parFor} in line~\ref{line:sortElemCompL}, \ref{line:sortElemCompR}
	& \makecell{Array $T$\\after line~\ref{line:sortCompForEnd}}
	& \makecell{\texttt{parFor}\\in line~\ref{line:sortRankStore}}\\ \Xhline{0.8pt}
 
	\multirow{44}{3pt}[-15pt]{\includegraphics[scale=0.9]{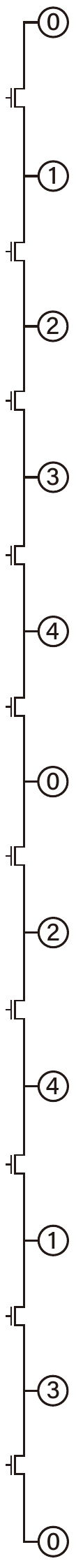}}
	&& \multirow{4}{*}{\makecell{\\$C_{0,0}$\\{\footnotesize$A[0]=8$}}}
	&&&
	& \multirow{4}{*}{\gape{\renewcommand{\arraystretch}{0.8}$T=\left[\color{gray}
		\begin{matrix} 0&\textcolor{black}{\textbf{1}}&0&1&1 \\ 
					   0&0&0&1&0 \\
					   1&1&0&1&1 \\ 
					   0&0&0&0&0 \\ 
					   0&1&0&1&0\end{matrix}\color{black}\right] $}}
	& \multirow{4}{*}{\Gape[4pt][0pt]{$R[0]=3$}}\\

	&&&&&&& \\ \cdashline{4-6} 

	&&&\multirow{2}{*}{$C_{1,0}$}
	&\multicolumn{1}{l|}{\Gape[4pt][0pt]{$i_r=1\nless i=0$}}
	&\multicolumn{1}{l|}{$A[1]=6 < A[0]=8$}
	&& \\

	&&&
	& \multicolumn{1}{r|}{\Gape[0pt][3pt]{$\therefore$ Receive $A[1]$ from right}}
	& \multicolumn{1}{r|}{\Gape[0pt][3pt]{$\therefore T[0][1]\leftarrow1$, Send 0 to right }} 
	&& \\ \cline{3-8}

	&& \multirow{4}{*}{\makecell{\\$C_{1,0}$\\{\footnotesize$A[1]=6$}}}
	& \multirow{2}{*}{$C_{0,0}$}
	& \multicolumn{1}{l|}{\Gape[4pt][0pt]{$i_l=0  <   i=1$}}
	& \multirow{2}{*}{\Gape[4pt][0pt]{Receive $0$ from left}}
	&  \multirow{4}{*}{\gape{\renewcommand{\arraystretch}{0.8}$T=\left[\color{gray}
		\begin{matrix} 0&1&0&1&1 \\ 
					   0&0&0&1&0 \\
					   1&1&0&1&1 \\ 
					   0&0&0&0&0 \\ 
					   0&1&0&1&0\end{matrix}\color{black}\right] $}}
	& \multirow{4}{*}{\Gape[4pt][0pt]{$R[1]=1$}}
	\\ 

	&&&
	& \multicolumn{1}{r|}{\Gape[0pt][3pt]{$\therefore$ Send $A[1]$ to left}}
	& 
	&& 
	\\ \cdashline{4-6}

	&& &\multirow{2}{*}{$C_{2,0}$}
	& \multicolumn{1}{l|}{\Gape[4pt][0pt]{$i_r=2\nless i=1$}}
	&\multicolumn{1}{l|}{$A[2]=9\nless A[1]=6$}
	&& 
	\\

	&&&
	& \multicolumn{1}{r|}{\Gape[0pt][3pt]{$\therefore$ Receive $A[2]$ from right}}
	&\multicolumn{1}{r|}{$\therefore$ Send $1$ to right} 
	&& \\ \cline{3-8} 

	&& \multirow{4}{*}{\makecell{\\$C_{2,0}$\\{\footnotesize$A[2]=9$}}}
	& \multirow{2}{*}{$C_{1,0}$}
	& \multicolumn{1}{l|}{\Gape[4pt][0pt]{$i_l=1  <   i=2$}}
	& \multicolumn{1}{l|}{Receive $1$ from left}
	&  \multirow{4}{*}{\gape{\renewcommand{\arraystretch}{0.8}$T=\left[\color{gray}
		\begin{matrix} 0&1&0&1&1 \\ 
					   0&0&0&1&0 \\
					   1&\textcolor{black}{\textbf{1}}&0&\textcolor{black}{\textbf{1}}&1 \\ 
					   0&0&0&0&0 \\ 
					   0&1&0&1&0\end{matrix}\color{black}\right] $}}
	& \multirow{4}{*}{\Gape[4pt][0pt]{$R[2]=4$}}\\%

	&&&
	& \multicolumn{1}{r|}{\Gape[0pt][3pt]{$\therefore$ Send $A[2]$ to left}}
	& \multicolumn{1}{r|}{$\therefore T[2][1]\leftarrow 1$}
	&& 
	\\ \cdashline{4-6}

	&& &\multirow{2}{*}{$C_{3,0}$}
	& \multicolumn{1}{l|}{\Gape[4pt][0pt]{$i_r=3\nless i=2$}}
	& \multicolumn{1}{l|}{$A[3]=5 < A[2]=9$}
	&& 
	\\

	&&&
	& \multicolumn{1}{r|}{\Gape[0pt][3pt]{$\therefore$ Receive $A[3]$ from right} }
	&\multicolumn{1}{r|}{$\therefore T[2][3]\leftarrow 1$, Send $0$ to right}
	&& \\ \cline{3-8} 

	&& \multirow{4}{*}{\makecell{\\$C_{3,0}$\\{\footnotesize$A[3]=5$}}}
	& \multirow{2}{*}{$C_{2,0}$}
	& \multicolumn{1}{l|}{\Gape[4pt][0pt]{$i_l=2<i=3$}}
	& \multirow{2}{*}{\Gape[4pt][0pt]{Receive $0$ from left}}
	&  \multirow{4}{*}{\gape{\renewcommand{\arraystretch}{0.8}$T=\left[\color{gray}
		\begin{matrix} 0&1&0&1&1 \\ 
					   0&0&0&1&0 \\
					   1&1&0&1&1 \\ 
					   0&0&0&0&0 \\ 
					   0&1&0&1&0\end{matrix}\color{black}\right] $}}
	& \multirow{4}{*}{\Gape[4pt][0pt]{$R[3]=0$}}\\

	&&&
	& \multicolumn{1}{r|}{\Gape[0pt][3pt]{$\therefore$ Send $A[3]$ to left} }
	& 
	&& 
	\\ \cdashline{4-6}

	&& &\multirow{2}{*}{$C_{4,0}$}
	& \multicolumn{1}{l|}{\Gape[4pt][0pt]{$i_r=4 \nless i=3$}}
	& \multicolumn{1}{l|}{\Gape[4pt][0pt]{$A[4]=7\nless A[3]=5$}}
	&& 
	\\

	&&&
	& \multicolumn{1}{r|}{\Gape[0pt][3pt]{$\therefore$ Receive $A[4]$ from right} }
	& \multicolumn{1}{r|}{$\therefore$ Send $1$ to right}
	&& \\ \cline{3-8} 

	&& \multirow{4}{*}{\makecell{\\$C_{4,0}$\\{\footnotesize$A[4]=7$}}}
	& \multirow{2}{*}{$C_{3,0}$}
	& \multicolumn{1}{l|}{\Gape[4pt][0pt]{$i_l=3<i=4$}}
	& \multicolumn{1}{l|}{Receive $1$ from left}
	&  \multirow{4}{*}{\gape{\renewcommand{\arraystretch}{0.8}$T=\left[\color{gray}
		\begin{matrix} 0&1&0&1&1 \\ 
					   0&0&0&1&0 \\
					   1&1&0&1&1 \\ 
					   0&0&0&0&0 \\ 
					   0&1&0&\textcolor{black}{\textbf{1}}&0\end{matrix}\color{black}\right] $}}
	& \multirow{4}{*}{\Gape[4pt][0pt]{$R[4]=2$}}\\

	&&&
	& \multicolumn{1}{r|}{\Gape[0pt][3pt]{$\therefore$ Send $A[4]$ to left}}
	& \multicolumn{1}{r|}{$\therefore T[4][3]\leftarrow1$}
	&& \\ \cdashline{4-6}

	&& &\multirow{2}{*}{$C_{0,1}$}
	& \multicolumn{1}{l|}{\Gape[4pt][0pt]{$i_r=0< i=4$}}
	& \multirow{2}{*}{\Gape[4pt][0pt]{Receive $0$ from right}}
	&& \\

	&&&
	& \multicolumn{1}{r|}{\Gape[0pt][3pt]{$\therefore$ Send $A[4]$ to right}}
	& 
	&& \\ \cline{3-8}

	&& \multirow{4}{*}{\makecell{\\$C_{0,1}$\\{\footnotesize$A[0]=8$}}}
	& \multirow{2}{*}{$C_{4,0}$}
	& \multicolumn{1}{l|}{\Gape[4pt][0pt]{$i_l=4 \nless i=0$}}
	& \multicolumn{1}{l|}{$A[4]=7< A[0]=8$}
	&  \multirow{4}{*}{\gape{\renewcommand{\arraystretch}{0.8}$T=\left[\color{gray}
		\begin{matrix} 0&1&0&1&\textcolor{black}{\textbf{1}} \\ 
					   0&0&0&1&0 \\
					   1&1&0&1&1 \\ 
					   0&0&0&0&0 \\ 
					   0&1&0&1&0\end{matrix}\color{black}\right] $}}
	&\\ 

	&&&
	& \multicolumn{1}{r|}{\Gape[0pt][3pt]{$\therefore$ Receive $A[4]$ from left} }
	& \multicolumn{1}{r|}{$\therefore T[0][4]\leftarrow1$, Send $0$ to left}
	&& \\ \cdashline{4-6}

	&& &\multirow{2}{*}{\Gape[4pt][0pt]{$C_{2,1}$}}
	& \multicolumn{1}{l|}{\Gape[4pt][0pt]{$i_r=2 \nless i=0$}}
	& \multicolumn{1}{l|}{$A[2]=9\nless A[0]=8$}
	&& \\

	&&&
	& \multicolumn{1}{r|}{\Gape[0pt][3pt]{$\therefore$ Receive $A[2]$ from right}}
	& \multicolumn{1}{r|}{$\therefore$ Send $1$ to right}
	&& \\ \cline{3-8} 

	&& \multirow{4}{*}{\makecell{\\$C_{2,1}$\\{\footnotesize$A[2]=9$}}}
	& \multirow{2}{*}{$C_{2,1}$}
	& \multicolumn{1}{l|}{\Gape[4pt][0pt]{$i_l=0< i=2$}}
	& \multicolumn{1}{l|}{Receive $1$ from left}
	&  \multirow{4}{*}{\gape{\renewcommand{\arraystretch}{0.8}$T=\left[\color{gray}
		\begin{matrix} 0&1&0&1&1 \\ 
					   0&0&0&1&0 \\
					   \textcolor{black}{\textbf{1}}&1&0&1&\textcolor{black}{\textbf{1}} \\ 
					   0&0&0&0&0 \\ 
					   0&1&0&1&0\end{matrix}\color{black}\right] $}}
	&\\ 

	&&&
	& \multicolumn{1}{r|}{\Gape[0pt][3pt]{$\therefore$ Send $A[2]$ to left} }
	& \multicolumn{1}{r|}{$\therefore T[2][0]\leftarrow1$ }
	&& \\ \cdashline{4-6}

	&& &\multirow{2}{*}{$C_{4,1}$}
	& \multicolumn{1}{l|}{\Gape[4pt][0pt]{$i_r=4\nless i=2$}}
	& \multicolumn{1}{l|}{$A[4]=7 < A[2]=9$}
	&& \\

	&&&
	& \multicolumn{1}{r|}{\Gape[0pt][3pt]{$\therefore$ Receive $A[4]$ from right}}
	& \multicolumn{1}{r|}{$\therefore T[2][4]\leftarrow 1$, Send $0$ to right}
	&& \\ \cline{3-8}

	&& \multirow{4}{*}{\makecell{\\$C_{4,1}$\\{\footnotesize$A[4]=7$}}}
	& \multirow{2}{*}{$C_{2,1}$}
	& \multicolumn{1}{l|}{\Gape[4pt][0pt]{$i_l=2< i=4$}}
	& \multirow{2}{*}{\Gape[4pt][0pt]{Receive $0$ from left}}
	&  \multirow{4}{*}{\gape{\renewcommand{\arraystretch}{0.8}$T=\left[\color{gray}
		\begin{matrix} 0&1&0&1&1 \\ 
					   0&0&0&1&0 \\
					   1&1&0&1&1 \\ 
					   0&0&0&0&0 \\ 
					   0&\textcolor{black}{\textbf{1}}&0&1&0\end{matrix}\color{black}\right] $}}
	&\\ 

	&&&
	& \multicolumn{1}{r|}{\Gape[0pt][3pt]{$\therefore$ Send $A[4]$ to left} }
	&
	&& \\ \cdashline{4-6}

	&& &\multirow{2}{*}{$C_{1,1}$}
	& \multicolumn{1}{l|}{\Gape[4pt][0pt]{$i_r=1< i=4$}}
	& \multicolumn{1}{l|}{\Gape[4pt][0pt]{Receive $1$ from right}}
	&& \\

	&&&
	& \multicolumn{1}{r|}{\Gape[0pt][3pt]{$\therefore$ Send $A[4]$ to right}}
	& \multicolumn{1}{r|}{$\therefore T[4][1]\leftarrow 1$}
	&& \\ \cline{3-8}

	&& \multirow{4}{*}{\makecell{\\$C_{1,1}$\\{\footnotesize$A[1]=6$}}}
	& \multirow{2}{*}{$C_{2,1}$}
	& \multicolumn{1}{l|}{\Gape[4pt][0pt]{$i_l=4\nless i=1$}}
	& \multicolumn{1}{l|}{$A[4]=7\nless A[1]=6$}
	&  \multirow{4}{*}{\gape{\renewcommand{\arraystretch}{0.8}$T=\left[\color{gray}
		\begin{matrix} 0&1&0&1&1 \\ 
					   0&0&0&\textcolor{black}{\textbf{1}}&0 \\
					   1&1&0&1&1 \\ 
					   0&0&0&0&0 \\ 
					   0&1&0&1&0\end{matrix}\color{black}\right] $}}
	&\\ 

	&&&
	& \multicolumn{1}{r|}{\Gape[0pt][3pt]{$\therefore$ Receive $A[4]$ from left} }
	& \multicolumn{1}{r|}{$\therefore$ Send $1$ to left}
	&& \\ \cdashline{4-6}

	&& &\multirow{2}{*}{$C_{3,1}$}
	& \multicolumn{1}{l|}{\Gape[4pt][0pt]{$i_r=3\nless i=1$}}
	& \multicolumn{1}{l|}{$A[3]=5 < A[1]=6$}
	&& \\

	&&&
	& \multicolumn{1}{r|}{\Gape[0pt][3pt]{$\therefore$ Receive $A[3]$ from right}}
	& \multicolumn{1}{r|}{$\therefore T[1][3]\leftarrow 1$, Send $0$ to right}
	&& \\ \cline{3-8}

	&& \multirow{4}{*}{\makecell{\\$C_{3,1}$\\{\footnotesize$A[3]=5$}}}
	& \multirow{2}{*}{$C_{1,1}$}
	& \multicolumn{1}{l|}{\Gape[4pt][0pt]{$i_l=1< i=3$}}
	& \multirow{2}{*}{\Gape[4pt][0pt]{Receive $0$ from left}}
	&  \multirow{4}{*}{\gape{\renewcommand{\arraystretch}{0.8}$T=\left[\color{gray}
		\begin{matrix} 0&1&0&1&1 \\ 
					   0&0&0&1&0 \\
					   1&1&0&1&1 \\ 
					   0&0&0&0&0 \\ 
					   0&1&0&1&0\end{matrix}\color{black}\right] $}}
	&\\ 

	&&&
	& \multicolumn{1}{r|}{\Gape[0pt][3pt]{$\therefore$ Send $A[3]$ to left} }
	&
	&& \\ \cdashline{4-6}

	&& &\multirow{2}{*}{$C_{0,2}$}
	& \multicolumn{1}{l|}{\Gape[4pt][0pt]{$i_r=0< i=3$}}
	& \multirow{2}{*}{\Gape[4pt][0pt]{Receive $0$ from right}}
	&& \\

	&&&
	& \multicolumn{1}{r|}{\Gape[0pt][3pt]{$\therefore$ Send $A[3]$ to right}}
	&
	&& \\ \cline{3-8}

	&& \multirow{4}{*}{\makecell{\\$C_{0,2}$\\{\footnotesize$A[0]=8$}}}
	& \multirow{2}{*}{$C_{3,1}$}
	&\multicolumn{1}{l|}{\Gape[4pt][0pt]{$i_l=3\nless i=0$}}
	& \multicolumn{1}{l|}{$A[3]=5 < A[0]=8$}
	&  \multirow{4}{*}{\renewcommand{\arraystretch}{0.8}$T=\left[\color{gray}
		\begin{matrix} 0&1&0&\textcolor{black}{\textbf{1}}&1 \\ 
					   0&0&0&1&0 \\
					   1&1&0&1&1 \\ 
					   0&0&0&0&0 \\ 
					   0&1&0&1&0\end{matrix}\color{black}\right] $}
	&\\ 

	&&&
	& \multicolumn{1}{r|}{\Gape[0pt][3pt]{$\therefore$ Receive $A[3]$ to left} }
	& \multicolumn{1}{r|}{$\therefore T[0][3]\leftarrow 1$, Send $0$ to left}
	&&\\ \cdashline{4-6}

	&&&&&\Gape[6pt][0pt]{}&& \\

	&&&&&\Gape[0pt][4pt]{}&& \\ \hline
 
	\end{tabular}
	\caption{Illustration of sorting $n=5$ elements using 1D-Crosspoint Array, with input data $A=\{8,6,9,5,7\}$. It is showing what is done in each \texttt{ParFor} loop.}
	\label{fig:5ex}
	\end{center}
	\vspace{-25pt}
\end{figure*}

\vspace{-5pt}\label{sec:parEnumSort}
In this section, we are going to show how the proposed 1D-Crosspoint Array can be used to carry out a parallel enumeration sort.
Sorting is a fundamental problem in data processing. Therefore, a fast sorting algorithm is important, especially for large sets of data. To this end, many parallel algorithms were introduced  with a myriad of  processor topologies. 
Here, we will show how our proposed topology sorts $n$ elements in $O(\lg n\lg\lg n)$ time, using $O(n^2)$ PEs. Furthermore, we will describe a way to reduce the time complexity to $O(1)$ using threshold logic gates and an encoder.
Among numerous algorithms and topologies, no $O(1)$ time sorting algorithm has been reported except those in \cite{Olariu1992mesh,Jang1992MeshColumnSort,Chen19943Dmesh}, where strong assumptions about path delays in a reconfigurable mesh were made. 
In particular, it is assumed that  a signal takes $O(1)$ time to travel through any path on such a topology, regardless of the distance between PEs. This assumption is based on the YUPPIE system described in~\cite{Maresca1989yuppie}. However,  in \cite{Maresca1989yuppie} Maresca et al., explain that the YUPPIE system needs a special clocking scheme to overcome the propagation delay, which is proportional to the distance between PEs. Another strong assumption would be that the overhead of inter-PE communications as compared to the amount of computation time used by PEs is negligible, but this defeats the purpose of having a mesh topology to interconnect PEs together. Unlike these  constant time algorithms, our proposed algorithm on 1D-Crosspoint Array does not need to make such a strong assumption as all communications take place between physically adjacent neighboring PEs.
\\The execution of the algorithm is illustrated in Figure~\ref{fig:4ex} for $n=4$ with an array $A=[6,7,8,5]$ of input values, and in Figure~\ref{fig:5ex} for $n=5$ with an array $A=[8,6,9,5,7]$ of input values. 
The proposed algorithm is a variant of enumeration sort. In general, enumeration sort has two tasks. First, it compares each element with every other element. It should be stated that we assume that the comparison of any two elements in $A$ takes $O(1)$ time. Otherwise, it should be factored into the overall time complexity of our sorting algorithm.  Second, when all comparisons are completed, for each element, it counts the number of elements that are less than that element, which then gives its rank.  The proposed algorithm follows this format as we now explain. 
\\The nested \texttt{parFor} loop in line~\ref{line:sortLoad} loads the $i$-th element of the input array $A$ to PE $C_{i,0}$ of every class $0\le i\le n-1$. It also initializes all elements of array $T$ to $0$. Each PE takes two steps to complete these two tasks. The latter task is completed in a single step as each PE $C_{i,0}$ has access to $T[i][\, ], 0\le i\le n-1$ without any contention by other PEs. Thus, PE $C_{i,0}$ can issue a master clear to all the $n$ flip-flops in row $i, 0\le i\le n-1.$ Alternatively, we can assign two flip-flops to each PE in class $i$ to clear a pair of flip-flops in two steps. In both cases, it takes $O(1)$ time to clear all the flip-flops in all of $T[i], 0\le i\le n-1.$  Loading the $i$-th element of $A$ to all PEs $C_{i,0}, 0\le i\le n-1$ in $O(1)$ steps is a little trickier. We assume that there is a backbone bus structure that facilitates this concurrent read operation by all the $n$ PEs. This can be viewed as a parallel load operation much like setting or clearing the bits of flip-flops in a register. The $i$-th element that is read from a memory where the $n$ elements of $A$ are stored is placed on a data bus,  which then becomes available to all PEs  $C_{i,0}, 0\le i\le n-1.$  Thus loading $A$ into $C_{i,0}, 0\le i\le n-1$ can be carried out in $O(1)$ time. Therefore, the execution of this \texttt{parFor} loop takes $O(1)$ time to complete overall.
\begin{algorithm}[t!]
	\small
	\begin{spacing}{1.15}
	\algrenewcommand\algorithmicindent{10pt}
	\algrenewcommand{\algorithmiccomment}[1]{// \textit{#1}}
	\newcommand{\LineComment}[1]{\hfill// \textit{#1}}
	\begin{algorithmic}[1]

	\Statex \Comment{$A$ is a vector which stores a set of $n$ elements}
	\Statex \Comment{$C_{i,j}$ represent $j$-th PE (or replicate) of class $i$,} 
	\Statex \Comment{where $0\leq i\leq n-1$, $0\leq j\leq \frac{n}{2}-1$.}
	\Statex \Comment{$C_{i_r,j_r}, C_{i_l,j_l}$ denote right and left neighbors of $C_{i,j}$.}
	\Statex \Comment{$T$ is an $n\times n$ matrix. }
	\Statex \Comment{$R$ is a vector which stores ranks of $n$ elements.}
	\vspace{3pt}
	\Statex \Comment{Load $A$ and initialize $T_i, 0\le i\le n-1$ in parallel}
	\ParFor{($0\leq i \leq n-1$)} \label{line:sortLoad}
		\State Load $A[i]$ to $C_{i,0}$
		\ParFor{$0\leq k\leq n-1$}
		    \State $T[i][k]\leftarrow 0$
		\EndParFor
	\EndParFor
	\Statex \Comment{Compare elements in parallel}
	\ParFor{$(0\leq i \leq n-1$, $0\leq j\leq \frac{n}{2}-1)$} \label{line:sortCompFor}
		\If{$(i_l < i)$} \label{line:sortCidCompL}
			\State Send $A[i]$ to left neighbor
			\If{$($Signal from left $== 1)$}
				\State $T[i][i_l]\leftarrow 1$
			\EndIf
		\Else
			\State Receive $A[i_l]$ from left neighbor
			\If{$(A[i_l] < A[i])~||~\big((A[i_l]=A[i])\&\&(i_l< i)\big)$}\label{line:sortElemCompL}
				\State $T[i][i_l]\leftarrow1$
				\State Send $0$ to left neighbor
			\ElsIf{$(A[i_l]>A[i])~||~\big((A[i_l]=A[i])\&\&(i_l\ge i)\big)$}\label{line:sortElemCompLelse}
				\State Send $1$ to left neighbor
			\EndIf
		\EndIf
		\If{$(i_r < i)$} \label{line:sortCidCompR}
			\State Send $A[i]$ to right neighbor
			\If{$($Signal from right $== 1)$}
				\State $T[i][i_r]\leftarrow 1$
			\EndIf
		\Else
			\State Receive $A[i_r]$ from right neighbor
			\If{$(A[i_r] < A[i])~||~\big((A[i_r]=A[i])\&\&(i_r< i)\big)$}\label{line:sortElemCompR}
				\State $T[i][i_r]\leftarrow1$
				\State Send $0$ to right neighbor
			\ElsIf{$(A[i_r]>A[i])~||~\big((A[i_r]=A[i])\&\&(i_r\ge i)\big)$}\label{line:sortElemCompRelse}
				\State Send $1$ to right neighbor
			\EndIf
		\EndIf
	\EndParFor \label{line:sortCompForEnd}
	\Statex \Comment{Store the rank}
	\ParFor{$0\leq i \leq n-1$,} \label{line:sortRankStore}
		\State  $R[i]=\sum_{k=0}^{n-1} T[i][k]$  
	\EndParFor
	\caption{sorting using a 1D-Crosspoint Array}
	\label{alg:sorting}
	\end{algorithmic}
	\end{spacing}
\end{algorithm}

\noindent The second \texttt{parFor} loop, starting at line~\ref{line:sortCompFor} performs a parallel enumeration sort, comparing every element with every other element. As the 1D-Crosspoint Array provides a crosspoint between any two PE classes,  this task is carried out in a distributed manner. Each pair of neighboring PEs essentially compare their input elements that they were loaded within the previous \texttt{parFor} loop as follows. Between any two neighboring PEs, the PE with the greater class number sends its element to the PE with the smaller class number (See the third column in Figures~\ref{fig:4ex} and \ref{fig:5ex}).
Then the PE with the smaller class number carries out the comparison between the received element and the element that it was assigned with. This is done by the \texttt{if-else} statements beginning in lines~\ref{line:sortCidCompL} and \ref{line:sortCidCompR}. 
The two \texttt{if-else} statements are identical except that one of them is for communicating with the righthand side PE, and the other is for communicating with the lefthand side PE.  
During each of these comparisons, if the PE that carried out the comparison has the greater element, it writes a $1$ to the $T[i][i_l]$ (or $T[i][i_r]$), then sends a $0$ to the neighboring PE with the smaller element to indicate that the comparison is done. 
When the PE that carries out the comparison has the smaller element, it sends a $1$ to the neighboring PE with the greater element, notifying it that $1$ has to be written to the $T[i_l][i]$ (or $T[i_r][i]$). At the end of this step, the array $T$ is set in such a way that $T[i][i_l]=1$ implies that $A[i_l]<A[i]$ (See, the fourth column in Figures~\ref{fig:4ex} and \ref{fig:5ex}).
It may appear to the reader that this step requires a memory that supports a concurrent write in order to run in $O(1)$ time.  
 \begin{figure}[t!]
 \centering{
\includegraphics[scale=0.42]{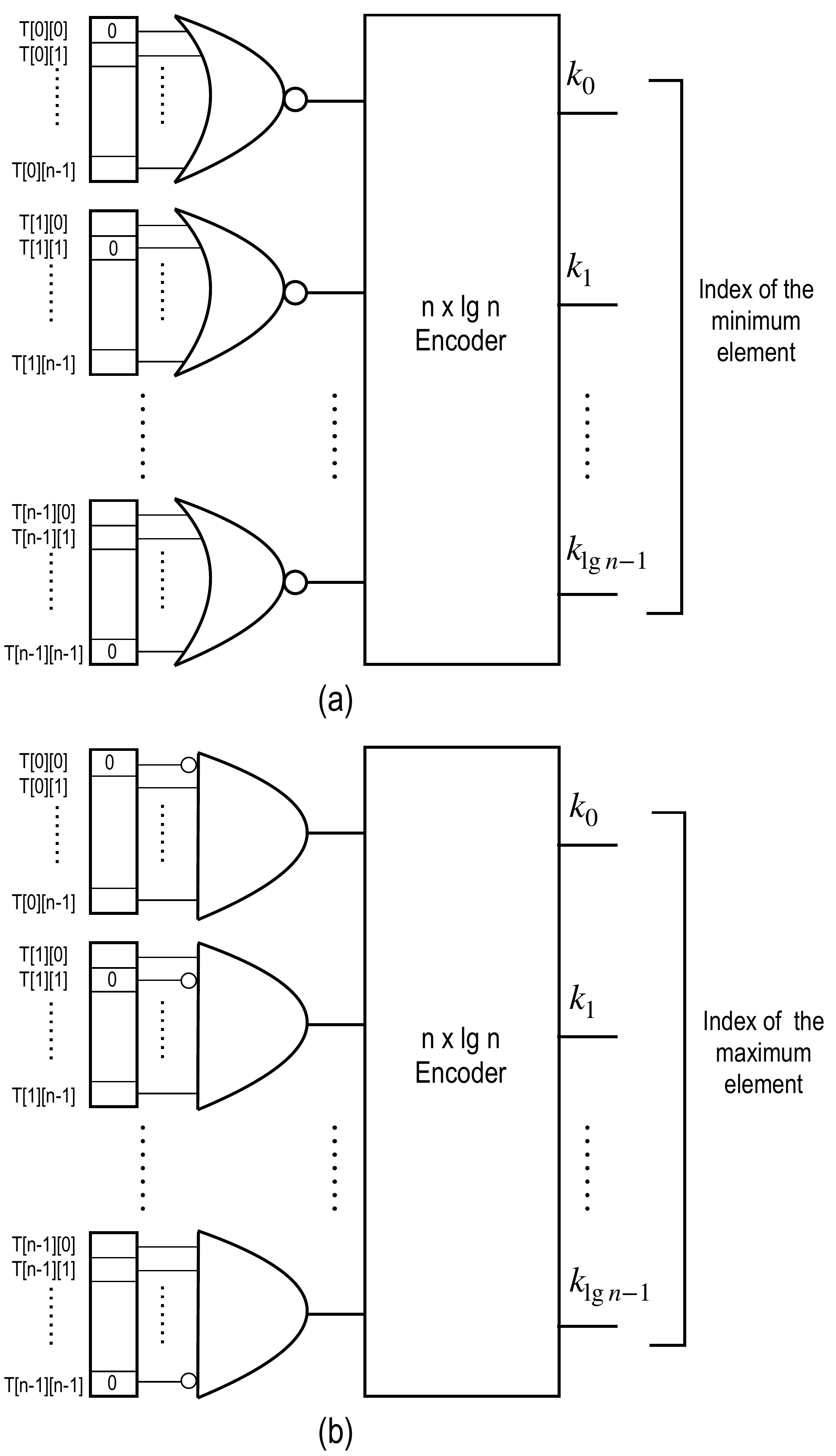}
\vspace{-5pt}
\caption{Computing the index of the minimum and maximum elements in an array of $n$ elements.}
\label{maxElement}
}
\end{figure}
However, we note that the \texttt{parFor} loop in line~\ref{line:sortCompFor} does not compare the same two elements more than once. 
By Remark~\ref{remk:oddNadj}, when $n$ is an odd number, the 1D-Crosspoint Array has only one adjacency between any two classes of PEs. Therefore, the locations to which the PEs write the comparison results are all distinct as  seen in Figure~\ref{fig:5ex}, and hence a concurrent write memory is avoided. 
On the other hand, in even $n$ case, by Remark~\ref{remk:evenNadj}, there are $\frac{n}{2}-1$ redundant adjacencies on the 1D-Crosspoint Array. For example, it is seen in Figure~\ref{fig:4ex} that the comparison between PE classes $1$ and $2$ is done twice, and therefore the PEs $C_{2,0}$ and $C_{2,1}$ within the same class of PEs, i.e., class 2,  attempt to write to location $T[2][1]$ at the same time (see the underlined elements in $T$ in the fifth column). To avoid multiple writes when $n$ is even, another PE can easily be added to switch to a 1D-Crosspoint Array with an odd number of PE classes. Thus, assuming that we have an odd number of PE classes, and given that  the number of \texttt{if-else} statements that a PE in any given class executes is at most two, this \texttt{parFor} loop also takes only $O(1)$ time to complete. 
Finally, the last \texttt{parFor} loop in line~\ref{line:sortRankStore} computes the  rank of each element, which corresponds to the second task of enumeration sort. This step is executed by each class of PEs separately, where up to $\frac{n}{2}$ PEs in each class are used to carry out addition operations. Thus, accessing the entries in $T[i][\, ]$ cannot result in any contention across the PE classes.  
We further note that each $1$ in $T[i][\, ]$ implies that there is an element less than $A[i],$ and hence counting  the number of $1$'s or summing the entries  in $T[i][\, ]$ gives the rank of the $A[i]$. In a way, the vectors stored in $T[i][\, ], 0\le i\le n-1$ may be viewed as encoded representations of the ranks of the elements in $A,$ and no further operations may be needed. This makes the time complexity of sorting a set of $n$ elements on an $O(n^2)$-PE-1D-Crosspoint Array $O(1)$ if the ranks of the elements in $A$ are not needed in decimal or some other, preferred number representation. 
\\Besides sorting, a significant utility of this encoded form of sorting is the ease with which the minimum or maximum of the elements in $A$ can be determined. Both these tasks can be completed in $O(1)$ time using simple logic gates, and an encoder consisting of OR gates with a fan-in of $n$. 
To see why, suppose that the index of the minimum element of $A$ is $i_{\rm min}$. Then $T[i_{\rm min}][\,]$ should be $[0,0,\cdots,0,0]$. Thus, if we apply a NOR function to the elements in each row as shown in Figure~\ref{maxElement}(a), only the row corresponding to the minimum element should output $1$. Therefore, we can compute $i_{min}$ by cascading $n$ $n$-input NOR gates with an $n$-input by $\lg n$-output encoder in $O(1)$ time. The time complexity increases to $O(\lg n)$ if the fan-in of OR gates within the encoder is assumed to be  a constant or  the fan-in of the NOR gate that is applied to each row of bits in $T$ is constant.  Similarly, suppose that the index of the maximum element of $A$ is $i_{max}$. Then $T[i_{\rm max}][]$ should be all $1'$s except  $T[i_{\rm max}][i_{max}]$. Therefore, if we apply an AND function to each row after complementing the diagonal elements in $T$, then only the row that corresponds to the maximum element should output $1$. Again, an encoder may be used to obtain the index of the maximum element in $A$ in $O(1)$ time with unlimited fan-in gates and in $O(\lg n)$ time with constant fan-in gates as shown in Figure~\ref{maxElement}(b). In both cases, we only use $O(\lg n)$ logic gates with $n$ fan-in for the encoder and $O(n)$ OR or AND gates for the first stage in Figure~\ref{maxElement}.
This querying procedure can be extended to answering other queries as well. For example, we can determine if any particular element in $A$ has a rank of $j$ or higher,  $1\le j\le n.$ The $j = 1$ case is trivial as it only requires an OR-gate with $n$ inputs. If $j=2$ then we can add the bits in the corresponding row in the $T$ matrix in pairs to determine if any particular element in $A$ has a rank of $j$ or higher with high probability  in $O(1)$ time. To see why, note that exactly three out of four binary patterns $00, 01, 10, 11$ result in a sum of zero or one, and hence the number of $n$-bit patterns in which all $n/2$ pairs sum to less than two is  $3^{n/2}.$ Dividing this by the total number of $n$-bit patterns, we find that the probability of incorrectly guessing that the rank of an element is at least two tends to 0 as the following computation shows 

\vspace{-5pt}
$$\frac{3^{n/2}}{2^n} = \frac{1}{2^{n(1-0.5\lg 3)}} = \frac{1}{2^{0.2175n}}\rightarrow 0 \text{ as } n\rightarrow \infty.\vspace{1pt}
$$
 
\vspace{3pt}\noindent
Thus $n/2$ 2-input AND gates together with an $n/2$-input OR gate suffice to determine if the rank of any given element is at least 2 with high probability.  In general, adding $k$ bits together, $k\ge 2$ leads to the probability of incorrectly guessing that the rank of an element is at least $j, 1\le j\le n$ is given by   

\vspace{-5pt}
$$\frac{\big\{2^k - \sum_{i=0}^{j-1} {k\choose i}\big\}^{n/k}}{2^n}  \rightarrow 0 \text{ as } n\rightarrow \infty.\vspace{1pt}
$$

\vspace{-3pt}\noindent
We note that it is always possible to determine the rank of any particular element in $A$ exactly, by summing the entries in the corresponding row in the $T$ matrix by a binary tree adder in $O(\lg\lg n)$ time. Querying the $i$-th largest element is only a little more involved.
Suppose that the index of the $i$-th largest element is $i_{x}.$ Then the sum of the 1's in $T[i_{x}][\,]$ must be equal to $i.$ Moreover, the sum of the 1's in no other row should be equal to $i_{x}$. Therefore, if we subtract $i$ from the sum of  all entries in the rows  in $T$ then the row that produces a zero provides the key to the $i$-th  largest element. Once identified, the index of this row can be captured as before using an $n$-input by $\lg n$-output encoder in $O(1)$ or $O(\lg n)$ time as in the prior cases. All we need to do is to flip the bits that  are generated by  subtraction operations, before feeding them to the encoder. The only other time complexity that must be added to the total time complexity is that of  a $\lg n$-bit subtraction operation. This can be carried out by the first $n$ PEs, $C_{i,0}, 0\le i\le n-1$ in $O(1)$ time, using $O(n\lg^2 n)$ logic gates with $O(n)$ fan-in or $O(\lg \lg n)$ gate-level time using a $\lg n$-bit prefix adder~\cite{brent1982adder} with $O(n\lg n)$ constant fan-in logic gates.
Another possible query would be  to search for a specific number in a given array. This can be accomplished with an encoder and slightly modified Algorithm~\ref{alg:sorting}. Since we only need to compare each element with the number we are searching for, not with $n-1$ other elements, only one replicate from each class is needed to check if the element it has is equal to the  number that is searched in the \texttt{if} statement in line~\ref{line:sortElemCompL}, and~\ref{line:sortElemCompR}. We would not need the following \texttt{else} blocks in line~\ref{line:sortElemCompLelse}, and~\ref{line:sortElemCompRelse}, and the array $T$ can be a vector of size $n\times 1$. Each element in array $T$ will then be fed into the $n\times \lg n$ encoder, which will output the index of the number that is  searched. Due to the fact that algorithm for search query is a cut-down version of the Algorithm~\ref{alg:sorting}, and does only require a $n\times \lg n$ encoder, the searching among $n$ elements take $O(1)$ time in the 1D-Crosspoint Array.
\begin{figure}[b!]
\centering{
\includegraphics[scale=0.39]{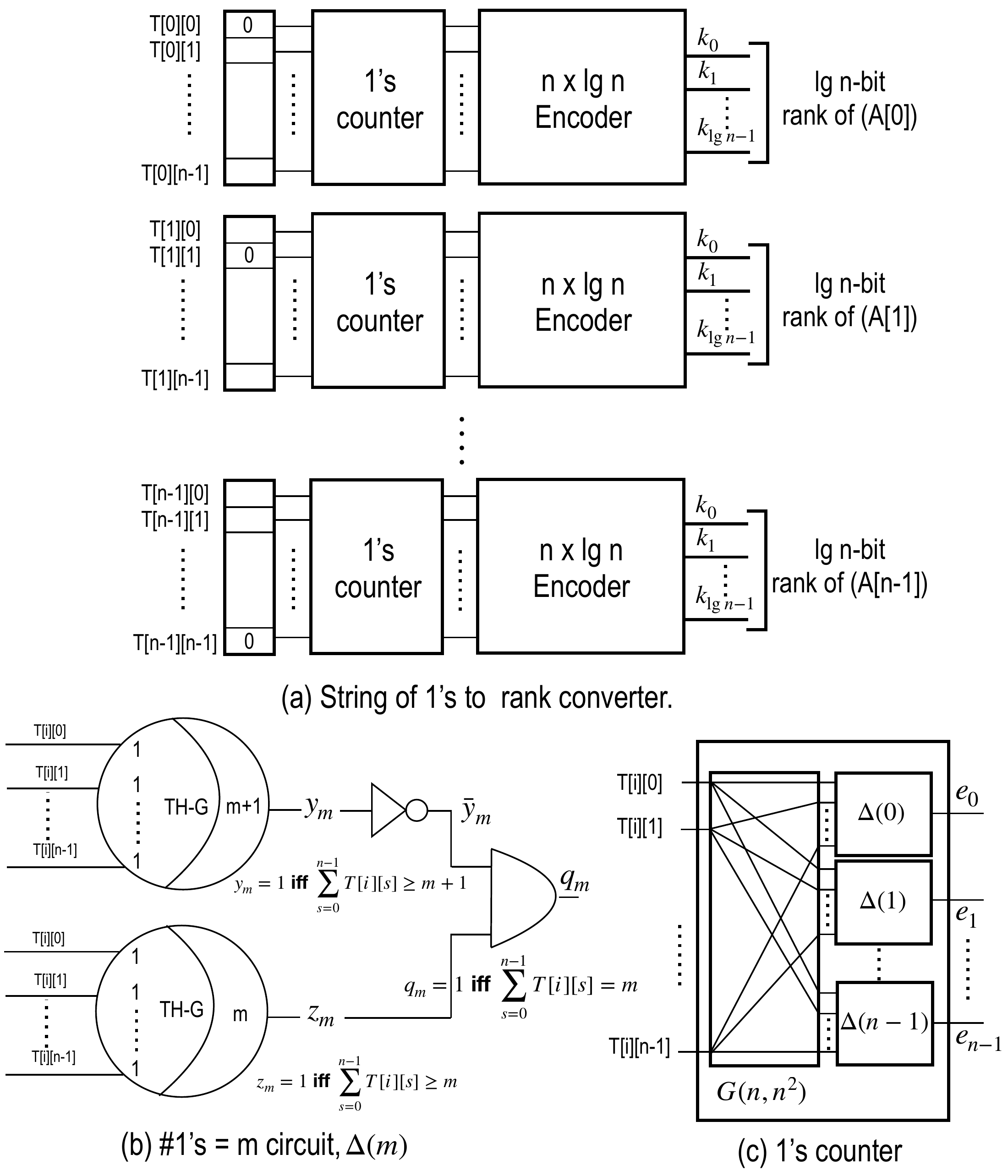}
\vspace{-5pt}
\caption{Computing the ranks from the $T$ matrices.}
\label{thresholdCircuit}
}
\end{figure}
\\Finally, if the ranks are desirable in decimal or some other numerical notation, the binary entries in $T[i][], 0\le i\le n-1$ can be summed in $O(\lg n\lg\lg n)$ time using a binary tree of $n$ Brent-Kung adders~\cite{brent1982adder}, each with two $\lg n$-bit operands and consisting of $O(\lg n)$ logic gates with constant fan-in to compute the ranks of all elements on $n$ PEs in parallel, where $C_{i,0}, 0\le i\le n-1$ computes the rank of $A[i].$ 
This increases the overall gate-level time complexity of sorting a set of $n$ elements from $O(1)$ to $O(\lg n\lg\lg n),$ using $O(n\lg n)$  additional logic gates with  constant fan-in per PE. We note that it is impossible to reduce the time complexity to $O(1)$ using  a polynomial order of  AND, OR and NOT gates even with unbounded fan-in~\cite{furst1984parity,parberry1988parallel,siu91,hajnal1993threshold}. 
Instead, we use threshold logic as shown in Figure~\ref{thresholdCircuit}. The diagram in Figure~\ref{thresholdCircuit}(a) depicts the overall layout of our architecture. The $i$th 1's-counter sets its $e_m$ output to 1 when its input string $T[i][]$ has exactly $m$ 1's, $0\le i\le n-1$. The $n$-input, $\lg n$-output encoder then generates the rank for $T[i][], 0\le i\le n-1.$ 
The 1's counters (part (c)) are constructed from  $\Delta(m)$ circuits, each of which is obtained by a pair of threshold gates, where  each such gate is assumed to have $O(1)$ delay, which is the standard assumption in threshold logic circuits, even though practical physical constraints may make this assumption overly optimistic~\cite{siu91}. 
The bipartite graph $G(n,n^2)$ replicates each of its $n$ inputs, and connects each copy to one of the inputs of the $n$ inputs of each $\Delta(m)$ circuit as shown in the figure. The upper threshold gate in a $\Delta(m)$ circuit, (in part (b)), together with an inverter produces a 1 when fewer than $m+1$ of its binary-valued inputs are equal to 1, whereas the lower threshold gate produces a 1 when $m$ or more of its binary-valued inputs are equal to 1. 
Combining the outputs of the two threshold gates with an AND gate thus detects if the number of 1's in the input string $T[i][]$ is equal to $m.$ Given that the path from an input to the output of a $\Delta(m)$ circuit traverses a threshold gate, an inverter, and AND gate, its output is computed in $O(1)$ time. Therefore, each output of a 1's counter is also computed in $O(1)$ time. 
\begin{table*}[b]
	\vspace{-8pt}
	\begin{center}
	\begin{tabular}{c?c?c?c}
	\Xhline{0.8pt}
	Architecture 		& \gape{\makecell{Hardware\\Complexity}}
	& Time Complexity 	& Comments \\ \Xhline{0.8pt}

	\Gape[6pt]{\makecell{Valiant's\\generic model}}
	& \makecell{when\\$4\le2n\le k\le \frac{n(n-1)}{2}$}
	& \gape{\makecell{$(\lg n-\lg\frac{k}{n})$\\$\cdot(\lg\lg n -\lg\lg\frac{k}{n})$}}
	& \gape{\makecell{Capable of arbitrary disjoint comparison\\without fan-in,fan-out restriction~\cite{valiant1975parallelism}.}}\\\hline

	\makecell{Preparata's\\algorithm}
	& $O(n \lg n)$ 
	& $O(\lg n)$
	& \Gape[3pt][1pt]{\makecell{A variant of enumeration sort that uses\\Valiant's sorting algorithm in the process~\cite{preparata1978mergeSort}. }}\\\hline

	\gape{\makecell{Bitonic Sorting\\Network}}
	& \gape{$O(n\lg^2n)$}
	& $O(\lg^2n)$
	& \gape{\makecell{Practical but its time complexity\\ grows faster than $O(\lg n)$~\cite{batcher1968bitonic}.}}\\\hline

	\gape{AKS Network}
	& \gape{$O(n\lg n)$}
	& $O(\lg n)$
	& \gape{\makecell{Very large constant\\ in time complexity~\cite{Ajitai1983AKS}.}}\\\hline

	\makecell{Reconfigurable\\Mesh}
	& $O(n^2)$
	& $O(1) $ 
	& \gape{\makecell{$n\times n$ mesh with switches at the crosspoints.\\Requires strong assumption\cite{Olariu1992mesh}\cite{Jang1992MeshColumnSort}.}}\\ \hline

	\makecell{1D-Crosspoint\\Array}
	& $O(n^2)$
	& \Gape[6pt]{\makecell{$O(\lg n(\lg\lg n))$, or\\$O(1)$ with threshold gates.}}
	& $n$ distinct PEs, and $O\left(\frac{n}{2}\right)$ replicates.\\ \Xhline{0.8pt}

	\end{tabular}
	\caption{Different parallel sorting architecture}
	\label{tab:sortCompare}
	\end{center}
\end{table*}
Given that the ranks are computed by a cascade of a 1's counter and an $n$-input, $\lg n$-output encoder, the ranks are computed in $O(1)$ time as well. Hence, sorting of $n$ numbers is completed in $O(1)$ time using $O(n^2)$ threshold gates. 
It should be pointed out that sorting is known to be in class $TC^0$ of problems~\cite{chandra1984constant} in which every problem can be solved using a Boolean circuit with $O(1)$ depth that is constructed out of AND, OR, and threshold gates that grows by a polynomial function of the problem size $n.$ Our rank computation architecture provides an actual circuit with threshold gates and other combinational circuit logic that complete our $O(1)$ time sorting algorithm with $O(n^2)$ gates, matching both the depth and circuit complexity bounds.
\vspace{-4pt}
\section{Discussion}
\label{sec:discussion}
\vspace{-5pt}
\noindent In this section, we compare and analyze our work with earlier results on comparison and sorting problems.  Table~\ref{tab:sortCompare} lists these results with their hardware and time complexities for sorting~\cite{valiant1975parallelism,preparata1978mergeSort,Ajitai1983AKS,batcher1968bitonic,Olariu1992mesh}. 
One such key contribution is due to Valiant who devised a general solution and its time complexity analysis for finding the maximum and sorting problems using a generic model of computation~\cite{valiant1975parallelism}.
The model assumes $k$ PEs that can compare arbitrary disjoint pairs at any given time. In addition, the PEs can share comparison results without any fan-in or fan-out restriction, or in other words, can send or receive comparison results without any time overhead.
Valiant showed that, for such  a model, when $4\le 2n \le k \le \frac{n(n-1)}{2}$, where $n$ is the problem size, the time complexity of finding a maximum is $\lg\lg n - \lg\lg\frac{k}{n}$(See \cite[p.350]{valiant1975parallelism}).
Despite the fact that our proposed 1D-Crosspoint Array requires an extra PE, the actual comparison is done on at most $\frac{n(n-1)}{2}$ PEs, and hence 1D-Crosspoint Array can be considered as an extreme case of Valiant's solution. Substituting $k=\frac{n(n-1)}{2}$, the time complexity of finding the maximum becomes $\lg\lg n-\lg\lg\frac{n-1}{2}\approx O(1)$. We
recall that the time complexity of finding the maximum on 1D-Crosspoint Array without fan-in restriction is also $O(1),$ which indicates that we have presented a tangible model of computation, rather than an abstract model with an impractical assumption about fan-in and fan-out of PEs, while matching Valiant's time complexity.
For sorting, Valiant proposed a parallel variation of merge sort that had an upper bound of $2\left(\lg n-\lg\frac{k}{n}\right)\left(\lg\lg n -\lg\lg\frac{k}{n}\right)$ for $4\le 2n \le k \le \frac{n(n-1)}{2}$ (See \cite[p.355]{valiant1975parallelism}). As in finding a maximum, he assumes unbounded fan-in and fan-out in this case as well, and this simplifies the time complexity of  
distributing workloads between consecutive stages of merging sublists by  $O(\lg k).$ 
Substituting $k=\frac{n(n-1)}{2}$ yields $2\left(\lg n-\lg\frac{n-1}{2}\right)\left(\lg\lg n -\lg\lg\frac{n-1}{2}\right)\approx O(1)$. On the other hand, 1D-Crosspoint Array has a sorting time complexity of $O((\lg n)\lg \lg n)$ if we restrict the fan-in of gates in our solution to $O(1)$. Without such a restriction, our 1D-Crosspoint Array together with $O(n^2)$ threshold gates and $n$-input, $\lg n$-output encoders sorts in $O(1)$ time as well. As mentioned in section~\ref{sec:parEnumSort}, the assumption of threshold gates having delay of $O(1)$ may be overly optimistic,
Valiant also admits that $O(\lg n)$ overhead remains to be overcome in sorting problems (See \cite[p.349]{valiant1975parallelism}).
However, here we have presented a specific architecture and a sorting algorithm that is well-suited to run on this architecture with the same time complexity as Valiant's parallel sorting algorithm when $k = n(n-1)/2.$
It should also be noted that the 1D-Crosspoint Array architecture is a computation model that has essentially no communication conflict or delay due to its structure being the simplest form of PE pairing. A more quantitative comparison of its performance with Valiant's parallel sorting algorithm is difficult as there are steps in his algorithm whose time complexities are not easily quantifiable.
Preparata utilized the Valiant's merging algorithm to devise a variation of enumeration sort where the Valiant's merge algorithm was used as a first step in his algorithm~\cite{preparata1978mergeSort}. Preparata's algorithm uses $O(n\lg n)$ number of PEs, and sorts $n$ elements in $O(\lg n)$ time. However, as it makes crucial use of Valiant's algorithm, it also follows the same assumptions and overlooks the $O(\lg k)$ time overhead in the merging step as well. 
\\\indent
In another direction, sorting networks provide an alternative to sorting algorithms.  A sorting network refers to  an architecture that is  composed of wires and comparators that can compare their inputs and sort them, where each element in the array to be sorted can be compared with one other element at a time. The sorting network proposed by Ajtai et al.~\cite{Ajitai1983AKS}, which is commonly known as  the AKS network named after its authors, has a hardware complexity of $O(n\lg n)$ and time complexity of $O(\lg n)$, which is smaller than the 1D-Crosspoint Array by a factor of $O(\lg\lg n),$ under the constant fan-in assumption. It  is also  faster than Valiant's parallel sorting algorithm by the same factor, when $k=n\lg n.$  It is worth noting that the AKS sorting network uses a more restrictive model even though it sorts faster than Valiant's algorithm. 
However, the constant involved in the $O(\lg n)$ term in~\cite{Ajitai1983AKS} equals 6100 in the lowest case~\cite{paterson1990improvedAKS}, which indicates that the 1D-Crosspoint Array(or Valiant's algorithm) will be faster for $\lg\lg n<6100$, or $n<2^{2^{6100}}$. In other words, for any practical $n$, 1D-Crosspoint Array will be faster in practice, even though the time complexity will grow faster then the AKS network. However, it should also be added that the AKS network uses $O(n\lg n)$ comparators as compared to $O(n^2)$ comparators in 1D-Crosspoint Array.
\\\indent Batcher's odd-even and  bitonic sorting networks~\cite{batcher1968bitonic} provide a more practical alternative to  parallel sorting algorithms such as those given by Valiant, Preparata and the one we presented here. Odd-even and bitonic sorting networks both  have a hardware complexity of $O(n\lg^2 n)$ and carry out sorting in $O(\lg^2 n)$ time complexity. The Valiant's upper bound on sorting with $k=n\lg^2 n$ PEs is $O(\lg n\lg\lg n)$, and this suggests that Batcher's sorting networks are not optimal. Nonetheless, it must be emphasized that Batcher sorting networks do not assume unbounded fan-in or fan-out in their comparators.  
\\\indent Besides sorting networks, a number of  reconfigurable array architectures have been used for sorting in the literature~\cite{lin1992sorting,Olariu1992mesh,Chen19943Dmesh,Jang1992MeshColumnSort,pan1998efficient,pan1998basic,pan1998linear,datta2002fast,he2009optimal}. An example we would like to mention here is the reconfigurable mesh architecture that Olariu proposed in~\cite{Olariu1992mesh}. This architecture has a hardware complexity of $O(n^2)$ and time complexity of $O(1)$. The  hardware complexity of this architecture matches the hardware complexity of 1D-Crosspoint Array and its time complexity matches the Valiant's upper bound. However, as explained in Section~\ref{sec:parEnumSort} of this paper, Olariu assumes that the path delays in the reconfigurable mesh are  $O(1).$ This is an impractical assumption, and if it is relaxed then the time complexity of the sorting algorithm on the reconfigurable mesh will be $O(n).$  

For future research, it will be worthwhile to obtain practical architectures with fewer than $O(n^2)$ processing elements to match the time complexity bounds for finding the minimum, maximum, searching and sorting problems under Valiant's parallel processor model.


\end{document}